\documentclass[article]{emulateapj}
\newcommand{\gsim}{\mbox{$\stackrel {>}{_{\sim}}$}} 
\newcommand{\lsim}{\mbox{$\stackrel {<}{_{\sim}}$}} 
 
\slugcomment{Accepted to ApJ}
 
\shorttitle{Resolved Chemistry in Maffei 2} 
\shortauthors{Meier \& Turner} 
 
\received{}

\begin{document} 
 
\title{Spatially Resolved Chemistry in Nearby Galaxies II. The Nuclear
Bar in Maffei 2}

\author{David S. Meier\altaffilmark{1,2}, and Jean L. Turner\altaffilmark{3}}  

\altaffiltext{1}{New Mexico Institute of Mining \& Technology, 801 Leroy Place, 
Socorro, NM 87801; dmeier@nmt.edu}
\altaffiltext{2}{National Radio Astronomy Observatory,
P. O. Box O, 1003 Lopezville Road, Socorro, NM 87801}
\altaffiltext{3}{Department of Physics and Astronomy, UCLA, Los
Angeles, CA 90095--1562; turner@astro.ucla.edu} 

\begin{abstract} 

We present 2$\arcsec$ - 10$\arcsec$ imaging of eleven transitions from 
nine molecular species across the nuclear bar in Maffei 2. The data were obtained with the BIMA
and OVRO interferometers.  The ten detected transitions are compared 
with existing CO isotopologues, HCN, CS and millimeter continuum data.  
Dramatic spatial variations among the mapped species are observed across 
the nuclear bar.  A principle component analysis is performed to 
characterize correlations between the transitions, star formation and 
molecular column density.  The analysis reveals that HCN, HNC, HCO$^{+}$ 
and 3 mm continuum are tightly correlated, indicating a direct connection 
to massive star formation. We find two main morphologically distinct chemical 
groups, CH$_{3}$OH, SiO and HNCO comprising the grain chemistry molecules, 
versus HCN, HNC, HCO$^{+}$ and C$_{2}$H, molecules strong in the 
presence of star formation.  The grain chemistry molecules, HNCO, CH$_{3}$OH 
and SiO, trace hydrodynamical bar shocks.  The near constancy of the HNCO/CH$_{3}$OH,
SiO/CH$_{3}$OH and SiO/HNCO ratios argue that shock properties are
uniform across the nucleus.  HCN/HCO$^{+}$, HCN/HNC, HCN/CS and HCN/CO 
ratios are explained primarily by variations in density. High 
HCO$^{+}$/N$_{2}$H$^{+}$ ratios are correlated with the C$_{2}$H line, 
suggesting that this ratio may be a powerful new dense photon-dominated 
region (PDR) probe in external galaxies. C$_{2}$H reveals a molecular outflow 
along the minor axis.  The morphology and kinematics of the outflow are 
consistent with an outflow age of 6-7 Myrs.  

\end{abstract}
 
\section{Introduction \label{intro}} 

Molecular abundances and excitation are sensitive probes of physical
processes taking place in molecular clouds \citep[e.g.,][]{VB98}.
Chemistry is pivotal both in controlling gas cooling and ionization
balance as well as in revealing changing cloud physical conditions
resulting from shocks and feedback from massive star formation and
galactic structure.  Millimeter telescopes are now capable of
surveying many molecular transitions in other galaxies, opening up
chemical studies to the rest of the Universe.

Single dish millimeter line surveys show that spectra of external
galaxies are extremely rich in molecular lines
\citep[e.g.][]{UGFMR04,MMMHG06,Cetal11,SNYHCIEL11,AMMMHOA11}.  For
nearby galaxies, giant molecular cloud (GMC) scale chemistry is within
reach of current millimeter interferometers.  In the first paper of
this series \citep[Meier \& Turner 2005; hereafter][]{MT05}, we
exploited this richness to carry out the first chemical imaging survey
of the nucleus of an external galaxy with spatial resolution high
enough to separate individual GMCs.  Images of eight astrochemically
important species were obtained in the nearby spiral IC 342, revealing
strong chemical differentiation among the GMCs.  A principal component
analysis demonstrated that the chemical variation is strongly
correlated with galactic structure such as bar and arm features, as
well as the presence and evolutionary state of star forming regions
\citep{MT05,MTS11}.

In this paper we extend the analysis to the nucleus of a second nearby
spiral galaxy, Maffei~2.  We have two main goals for this survey.  The
first is to test the generalizability of the results discovered in
\citet[][]{MT05}.  Maffei 2 and IC 342 are at the same distance and
their nuclear morphologies are similar.  Both have two molecular
spiral arms terminating on a central ring, with the central ring being
the site of intense, young, massive star formation.  It is reasonable
to expect that the two nuclei should exhibit similar chemistries if
the results of \citet[][]{MT05} for IC~342 are applicable to barred
spiral nuclei in general.

The second goal is to investigate the chemistry of Maffei 2 with
particular attention focused on its unique features.  \citet[][]{MT05}
find that in IC~342, shocks appear to be an important influence on the
nuclear chemistry and speculate that the shocks are associated with
orbital resonances in a bar potential.  The bar in IC~342 is weak
\citep{CTH+01}; however, and there remain other explanations for the
nuclear morphology \citep[][]{SBMC08}.  It is important to follow up
with an object that has a well established, strong bar.  Kinematic
modeling leaves little doubt that the nucleus of Maffei 2 hosts a
strong, highly defined nuclear bar of 320 pc radius \citep[Meier et
al. 2008, hereafter][]{MTH08,KNSS08}.  Because of the strength of the
bar, we expect that shock chemistry should be more important than in
IC~342.  Maffei 2 also has a somewhat more intense and extended starburst than
IC~342 \citep[][]{TH83,TH94}, so photon-dominated region (PDR) effects
may be more prominent than they are in IC~342.

\section{Observations \label{obs}} 

To pursue these goals, we obtain aperture synthesis observations 
of  eleven important molecular transitions
over the central R$\sim$0.5 kpc of the nuclear bar in Maffei 2, for
comparison with CO isotopologues \cite[][]{HT91,MTH08}, HCN(1--0)
\citep[][]{MTH08} and CS \citep[][]{KNSS08}.  Table \ref{MolP} lists 
molecular parameters for each 
transition.  With the Owens Valley Radio Observatory (OVRO) millimeter 
interferometer we observed three lines, HCO$^{+}$(1--0), 
HNCO(5$_{05}$--4$_{04}$) and SiO(1--0; $v$ =1).  HCO$^{+}$(1--0) 
and SiO(1--0; $v$ = 1) were observed between 1999
January 28 and 1999 March 29 along with the HCN(1--0) dataset
\citep{MTH08}.  HNCO(5$_{05}$--4$_{04}$) was observed along with
$^{13}$CO(1--0) and C$^{18}$O(1--0) between 1998 October 19 and 1999
January 5 (Table \ref{ObsT}).  The OVRO interferometer consisted of  
six 10.4 meter antennas equipped with SIS receivers \citep{OVRO91,
OVRO94}.  Transitions were observed in L, H and UH array
configurations except for the HNCO, which was observed in L and H,
providing spatial resolutions of $\sim2.^{''}5$ (Table \ref{ObsT}).
Data were calibrated using the in-house OVRO calibration package, MMA. 
Phase calibration was done by observing the point source 0224+671 every 
20 minutes.  Absolute flux calibration was based on observations of Neptune or
Uranus with 3C 273, 3C 84 and 3C 454.3 as supplementary flux
calibrators.  Based on the derived fluxes and flux histories of these
secondary flux calibrators we estimate that the absolute fluxes are
good to 10 - 15\%.  Two pointings were mosaiced to cover the nuclear
bar and therefore maps are corrected from the primary beam response.

The eight remaining transitions were observed with the ten element
Berkeley - Illinois - Maryland Association (BIMA)\footnote{Operated 
by the University of California, Berkeley, the University of
Illinois, and the University of Maryland with support from the
National Science Foundation.}  array \citep[][]{BIMA96} between 2002
September 08 and 2003 March 18 (Table \ref{ObsT}).  The eight spectral
lines were observed in two sets of spectrometer configurations.
C$_{2}$H(N = 1 -- 0, J = $\frac{3}{2}$--$\frac{1}{2}$), C$_{2}$H(N = 1
-- 0, J = $\frac{1}{2}$--$\frac{1}{2}$), SiO(2--1;$v=0$), HNC(1--0)
and HC$_{3}$N(10--9) were observed together, as were
CH$_{3}$OH($2_{k}-1_{k}$), C$^{34}$S(2--1) and N$_{2}$H$^{+}$(1--0).
Data were obtained in B, C and D configurations for all transitions
giving beamsizes of $\sim 7^{''} - 10^{''}$ depending on frequency and
sensitivity.  Calibration and data reduction was done in MIRIAD.
Phase calibration was also done by observing the same point source 0224+671
(0228+673) interlaced with the source every 24 minutes.  Absolute flux
calibration was done by observing W 3(OH).  Based on the derived
fluxes and flux histories of 0224+671 we estimate that the uncertainties 
in absolute flux are 10 - 15\%, similar to the OVRO data.  The $\sim 
130^{''}$ FWHM primary beam of the BIMA antenna is large enough such
that only one pointing was required. No corrections for primary beam
attenuation have been applied.  

Image analysis was done with the NRAO AIPS package.  
In making the integrated intensity maps emission greater than 1.2$\sigma$ 
was included.  Continuum emission has not been subtracted from the maps 
since the 3 mm continuum peak is below 1$\sigma$.  

\section{Results}

\subsection{A Sketch of the Nuclear Region of Maffei 2 \label{morph_gen}}  

\begin{figure}
\figurenum{1}
\epsscale{0.95}
\plotone{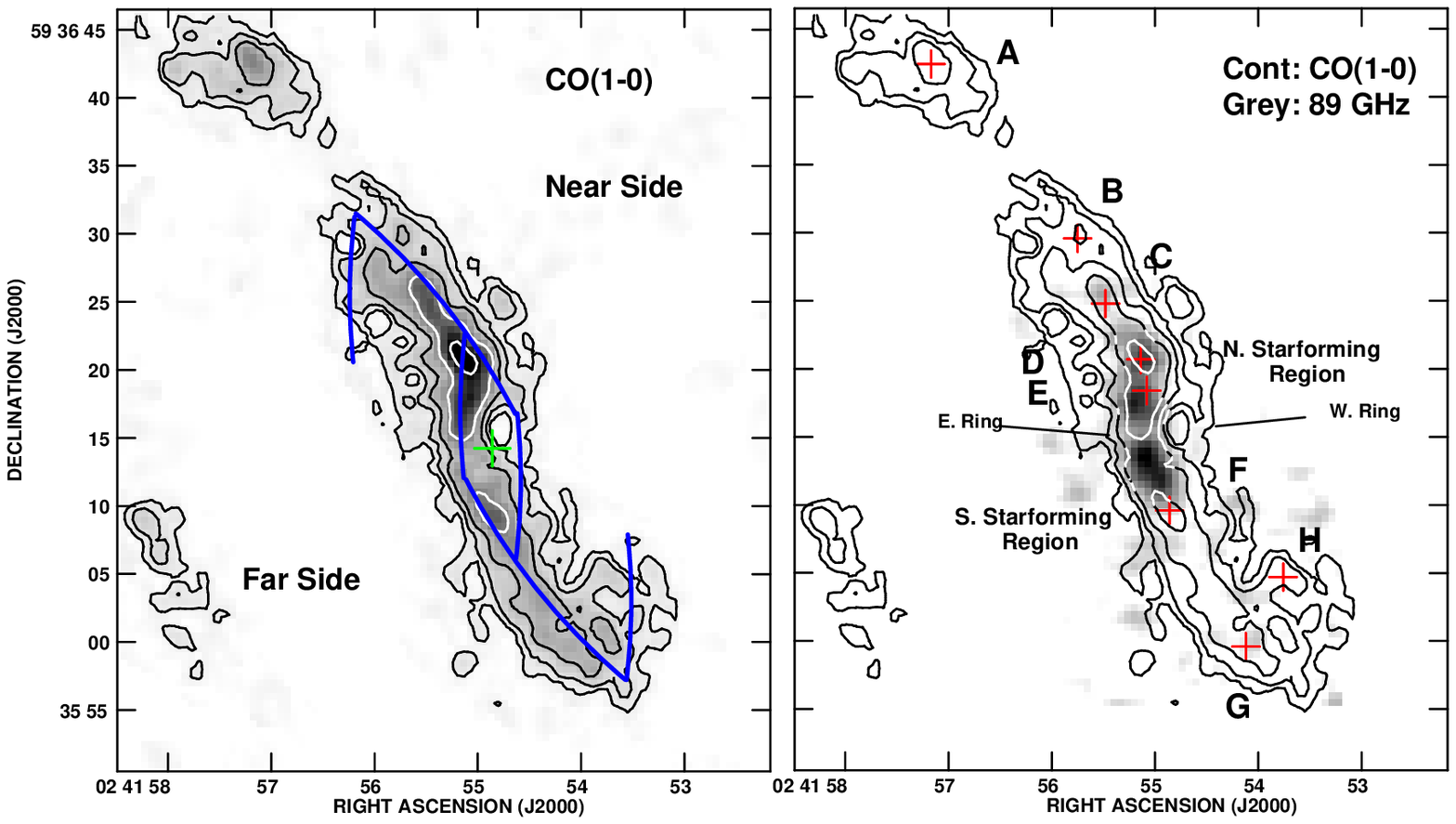}
\caption{The basic structure of Maffei 2's nucleus.  {\it Left)} The high resolution $^{12}$CO(1--0) 
integrated intensity map from \citet[][]{MTH08} with the location of the bar arms shown (in blue: 
online edition). Contours are 5.0 Jy bm$^{-1}$ km s$^{-1}$ $\times$ 1, 2, 4, 8, 16. A green cross 
marks the dynamical center.  {\it Right)} The same CO contours overlaid on a greyscale of the 
89 GHz continuum \citep[][]{MTH08}. The greyscale ranges from 0.5 - 5.0 mJy bm$^{-1}$. The 
red crosses represent the location of the labeled GMCs.  The location of the central ring and 
active star formation complexes are also noted. \label{Fintro} }
\end{figure}

The Sb galaxy Maffei~2 is one of the nearest and brightest extragalactic molecular
line sources, at a distance of only 3.3 Mpc \citep[1\arcsec~$\sim$15~pc,]
[see discussion in MTH08]{FLMR07}. A  strongly barred and somewhat 
asymmetric galaxy, Maffei~2 has undergone
a recent interaction with a small companion that has driven a large
quantity of gas into the nucleus, building up a compact central bulge
seen in the NIR \citep[][]{HMGT93,HTH96}.  This bar-like compact
bulge, has further driven the molecular gas into a central molecular zone of 
radius $R\sim 300$~pc that is barred and highly inclined 
\citep[][]{Ietal89,HT91}.  This nuclear bar, containing M(H$_2$)$\sim 
2\times 10^7~\rm M_\odot$ of gas \citep[][]{MTH08}, is structurally
reminiscent of a miniature version of large scale gaseous bars
\citep[][]{A92} (Fig. \ref{Fintro}).  Using 2 cm and 3 mm radio / mm continuum to trace ionizing 
radiation it is seen that the intersections of bar arms and central ring are sites of 
intense localized star formation \citep[SFR $\sim$ 0.04 M$_{\odot}$
yr$^{-1}$ per GMC or a total of $\rm L_{OB}\sim
10^8~L_\odot$;][]{TH94,TTBCHM06,MTH08}.  Averaged over the 
inner $\sim20^{''}$ radius, gas consumption timescales decrease below 
100 Myrs, making Maffei 2's nucleus a starburst.  

A central molecular ring of radius $\sim 7$\arcsec\ 
(~110~pc) is clearly resolved in gas kinematics, although the ring is
less apparent in integrated intensity images \citep{MTH08}.  GMCs D,
E, and F (cloud designations following MTH08 are shown in
Figs.~\ref{Fintro}) are located along the eastern side of the nuclear
ring.  Bar arms (including GMCs B and C in the north and G and H in
the south) extend off the leading edges of the ring. The bright GMCs
of the central ring, typically $\sim$60--80~pc in extent, and with
masses of a few $10^6~\rm M_\odot$, delineate the sites where gas
flowing in along the bar arms piles up at the inner ring.  Dense gas traced by HCN preferentially 
picks out the arm-ring intersections.   Gas not incorporated into the
dense component at the starburst locations is tidally sheared into a
moderate density, smoothly distributed ring  \citep{MTH08}. Massive star formation is
largely absent in the western side of the ring \citep[][]{MTH08}. Based on HCN, 
the fraction of dense gas is significantly lower towards the bar
ends.  Densities derived from CO tend to be fairly constant
(n$_{H2}~\sim 10^{3}$) over the inner bar, with the clouds associated with 
young star formation slightly warmer
($T_K\sim$ 30 -- 40 K) than the others (T$_{k}~ \sim 20$ K).  Gas
clouds on the quiescent western side of the nuclear ring are slightly cooler
yet \citep[][]{MTH08}.

\subsection{Molecular Abundances in Maffei 2\label{abuintro}}

Column densities are determined assuming optically thin
emission, and LTE:
\begin{equation}
N_{mol}~=~ \left(\frac{3k Qe^{E_{u}/kT_{ex}}}{8\pi^{3}\nu S_{ul}\mu_{0}^{2} 
g_{K_{u}}g_{I_{u}}} \right)I_{mol},
\end{equation}
where $S_{ul}$, $g$ and $E_{u}$ are the line strength, degeneracy and
upper energy of each state, respectively, and $\rm T_{ex}$ is the
excitation temperature associated with the transition.  Given that we
have mapped only one transition of each species, corrections for
background radiation and opacity have been ignored.  Column densities
are sensitive to $\rm T_{ex}$ through the partition function, $Q$, and
the energy of the upper state.  The asymmetric tops (HNCO and
CH$_{3}$OH) are more sensitive to temperature changes than are the linear
rotors.  Changes in gas density also affect excitation, particularly
for molecules with high critical densities (HNC, N$_{2}$H$^{+}$ and
HC$_{3}$N).  Densities have been at least
partially constrained by observation in Maffei 2 \citep{MTH08}.  
We adopt a T$_{ex}$ = 10 K for the molecular transitions here.  This
is similar to what is derived from the $^{13}$CO data
\citep[][]{MTH08} and is the same as assumed for IC~342
\citep[][]{MT05}, permitting easy comparison.  However T$_{ex}$ = 10 K 
may slightly underestimate excitation at the massive star forming regions.   Measured line intensities, 
peak temperatures, line centroids and line widths for the GMCs are 
reported in Table \ref{IntT}.  Fractional abundances ($X$$\rm(mol)\equiv 
N_{mol}/N_{H_{2}}$) calculated from the intensities given the above assumptions 
are listed in Table \ref{AbuT}, based on molecular 
parameters in Table \ref{MolP}. 

Fractional abundances also require an $\rm H_2$ column density,
N(H$_{2}$).  N(H$_{2}$) is most easily obtained from the CO(1-0)
brightness and an empirical Galactic conversion factor, $\rm X_{CO}$.
However, $\rm X_{CO}$ overpredicts N(H$_{2}$) in nearby galaxy
centers, including Maffei 2, by factors of a few \citep[][]{MTH08}.
An alternative measure of N(H$_{2}$) can be obtained from the CO
isotopologues, which tend to be optically thin.  We use the
$^{13}$CO(1-0) integrated intensity and [H$_{2}$/$^{13}$CO] = $7.0
\times 10^{5}$ ([$^{12}$CO/$^{13}$O] = 60 and [CO/H$_{2}$] =
$8.5\times 10^{-5}$) when calculating the H$_{2}$ column densities
\citep[see][for a detailed discussion of N(H$_{2}$) and its
uncertainties in Maffei 2]{MTH08}.

Fractional abundances for the species below are estimated to be
uncertain to at least a factor of three, although the relative column
densities---that is, the relative spatial distributions within the
nucleus of the different molecules---should be more
reliable. \citet[][]{MT05} has a detailed discussion of the
uncertainties associated with the abundance estimates.  Table
\ref{MolP} records for reference, the change in derived column
densities if T$_{ex}$ is changed from 10 K to 50 K, more typical of
the large beam gas {\it kinetic} temperatures.

\subsection{The Molecules of Maffei~2\label{morph_spec}}

We now discuss the morphologies of each species. Spectra taken at selected
locations across the nucleus with 8$\arcsec$ apertures are shown for
each transition in Fig. \ref{Fspec}.
Fig. \ref{FintiO} displays maps of the transitions detected with OVRO and
Fig. \ref{FintiB} displays those detected with BIMA. In 
the subsequent analysis we also include 3mm continuum emission and 
CO isotopologues from \citet[][]{MTH08}. The millimeter continuum in 
Maffei~2 is dominated by free-free emission rather than dust emission \citep[][]{MTH08}. 

\noindent{\it SiO \label{morph_sio}---Silicon Monoxide:} The J = 2--1
rotational line of both the $v$ = 0 and 1 vibrational states
were observed.  The $v$ = 1, J = 2 -- 1 transition is a maser in the
Galaxy \citep[][]{SB75}, but we do not detect it in Maffei~2.
The thermal $v$ = 0 transition of SiO is tentatively detected toward
GMCs B, F and G with peak temperatures over the 8$\arcsec$ aperture of
35 mK (Fig. \ref{FintiB}).  The integrated intensity map is noisy but
consistent in morphology to HNCO and CH$_{3}$OH, extended
along the southern arm.  Implied SiO abundances reach $\sim 1\times
10^{-10}$.  This abundance is lower than estimated for IC 342 and the nearby
starburst nuclei, NGC 253 and M 82 
\citep[][]{GMFN00,GMFN01,UGMFN06}, but typical of diffuse Galactic
clouds \citep[][]{GON96,T98sio} away from molecular outflows
\citep[][]{MBF92}; it is much larger than what is found for Galactic
dark clouds \citep[e.g.,][]{ZFI89}.

\noindent{\it C$_{2}$H \label{morph_c2h}---Ethynyl:} We detect both
the J = $\frac{3}{2}-\frac{1}{2}$ and $\frac{1}{2}-\frac{1}{2}$ fine
structure components of the N = 1--0 line.  C$_{2}$H is
brightest towards the northern star forming ring (GMC D+E), peaking at
0.11 K, and weakens significantly along the bar arms
(Fig. \ref{FintiB}).  Towards GMCs F and G, C$_{2}$H avoids the
column density peaks traced in $^{13}$CO(1--0).  Line widths are
significantly broader than those seen in the other weaker lines, so
the hyperfine structure of each fine structure component is being
marginally resolved here.  Evidence for the most blueshifted N = 1 --
0, J = $\frac{1}{2}-\frac{1}{2}$, F = 1 -- 0 hyperfine component is
seen towards the brightest GMCs (see Fig. \ref{Fspec}).  There is a plume
of emission along the minor axis in the $\frac{3}{2}-\frac{1}{2}$
line. From the relative intensities of the fine structure lines we can
determine the C$_{2}$H opacity.  The ratio of the brightness
temperature of the two fine structure transitions,
I($\frac{3}{2}-\frac{1}{2}$)/I($\frac{1}{2}-\frac{1}{2}$) ranges
from 1.2 - $\sim$2.8, with lowest values toward the regions with the
brighter emission.  For optically thin, LTE excitation a 
this ratio should be 2.0. Therefore C$_{2}$H opacities are significant along the
central ring, with implied values for the $\frac{3}{2}-\frac{1}{2}$
transition of $\tau_{CCH} ~\simeq$ 1.5.  Along the arms $\tau_{CCH}
<$1 except at  regions around GMC G away from the $^{13}$CO(1--0) 
column density peaks where
$\tau_{CCH} \sim 3$. Hence, despite its lower signal-to-noise, the
$\frac{1}{2}-\frac{1}{2}$ transition probably represents a closer
picture of the true C$_{2}$H abundance distribution.  Fractional
abundances assuming C$_{2}$H is optically thin peak at $\sim 2.5\times
10^{-8}$.  If we account for opacity, the abundance rises to $\sim
4.8\times 10^{-8}$.  These are similar to those seen towards the PDR
site in IC 342 \citep[][]{MT05} and at the high end of the range seen
in Galactic cores \citep*[][]{WBGLS80,HCK84,W83,TTH99,LL00}.

\noindent{\it HCO$^{+}$ \label{morph_hco}---Formyl Ion:} 
HCO$^{+}$(1--0) is the brightest line of this sample, and is 
imaged at high resolution ($\sim 2^{''}$).  The emission is dominated 
by GMCs D+E and F, with line peaks of 2.0 K (0.58 K over the $8^{''}$ 
aperture) (Fig. \ref{FintiO}).  Like HCN(1--0) \citep[][]{MTH08},
HCO$^{+}$(1--0) is primarily confined to the central ring region.
There is a second blueshifted velocity component at $\sim$-135 km
s$^{-1}$ (LSR) along the northern portion of the bar.  HCO$^{+}$(1--0)
abundances in Maffei 2 are between 2 - 8$\times 10^{-9}$, but these
(along with those of HCN and HNC) should be considered lower limits
since these very bright transitions may have significant optical
depths (see section \ref{hcohcn}).  These agree well with the
HCO$^{+}$ abundances in most Galactic dense clouds
\citep[eg.][]{PZLMS95,LL96,T95} and single-dish extragalactic
measurements \citep[][]{NJHTM92}.

\noindent{\it HNC \label{morph_hnc}---Hydrogen Isocyanide:} The J =
1--0 line of the linear molecule HNC is bright over the whole nuclear
bar, with peak temperatures $\sim$0.25 K (Fig. \ref{FintiB}).  At
this resolution, the morphology matches closely both $^{13}$CO(1--0)
and HCN(1--0), being dominated by the bright clouds, D + E and F, but
HNC emission is seen from every cloud including the non-nuclear bar
GMC, A.  Towards the northern side of the bar HNC emission is
especially bright at higher blueshifted velocities, similar to the
component seen in HCO$^{+}$(1--0).  HNC abundances are $1 - 2 \times
10^{-9}$, in agreement with IC 342 \citep[][]{MT05}.  These abundances
are intermediate between typical massive star forming cores and
Galactic dark clouds \citep[][]{WESV78,BSMP87,HYMO98,Num00}.

\noindent{\it HC$_{3}$N \label{morph_hc3n}---Cyanoacetylene:} We
mapped the J = 10--9 rotational line of HC$_{3}$N.  This molecule has
the largest electric dipole moment and the highest upper energy state
of the (thermal) sample (Table \ref{MolP}).  It is thus not surprising
that we observe HC$_{3}$N to be faint and spatially confined
(Fig. \ref{FintiB}).  To improve S/N, the HC$_{3}$N map was 
smoothed to a slightly larger beam size, but even then peaks at only 
$\sim$ 50 mK towards two unresolved locations at the
intersection of the ring and bar arms (C and F).  Abundances are $\sim 10^{-9}$,
similar to those in IC 342 \citep[][]{MT05,MTS11}, somewhat higher
than in cold Galactic clouds but much lower than the localized
Galactic hot core values \citep[$>10^{-7}$;
eg.][]{MTPZ76,VLSW83,COM91,DMNC00}.

\noindent{\it N$_{2}$H$^{+}$ \label{morph_n2h}---Diazenylium:} In the
Galaxy, N$_{2}$H$^{+}$ is regarded as a tracer of quiescent gas. We
observed the J = 1--0 rotational state of this linear ion, which has 
hyperfine splitting much smaller than the observed linewidths.  As is the 
case in IC 342, N$_{2}$H$^{+}$ is bright and extended across the nucleus 
(Fig. \ref{FintiB}).  Emission peaks towards GMC F with a peak brightness 
temperature reaching $\sim$0.15 K.  GMCs D and E, the closest 
clouds to the northern star formation region, show weak emission in
N$_{2}$H$^{+}$: $X(\rm N_{2}H^{+})$ reaches $\sim 7\times 10^{-10}$ 
at GMC F and drops by nearly a factor of three towards the 3 mm continuum 
peak.  The highest abundances are similar to what is seen in IC~342 and 
dense, quiescent cores in the Galactic disk \citep[][]{WZW92,BCM98}.

\noindent{\it C$^{34}$S \label{morph_c34s}---Carbon Monosulfide:} This
is an isotopologue of the dense gas tracer CS. The J = 2--1 rotational
transition of C$^{34}$S is tentatively detected only towards GMCs A, C
and F, at $< 30$mK (Fig. \ref{FintiB}).  The tentative detections imply 
$X(\rm C^{34}S) $ abundances of at most a few $\times 10^{-10}$, typical 
of Galactic dense cores and diffuse/translucent clouds \citep[for a
C$^{32}$S/C$^{34}$S isotopic abundance of 23; 
eg.][]{N84,DKV89,Wang93,CHWLC96,MELP97,LSJZ98,LL02}.  In the GMCs
near the northern star forming region abundances are an order of magnitude lower.  Evidently
C$^{34}$S(2--1) is not strongly enhanced anywhere across the nucleus,
and especially not at the massive star formation sites.  Limits on the CS/C$^{34}$S
isotopologue ratio assuming optically thin CS ranges from 4 to $>$14
based on the CS(2-1) line of \citet[][]{KNSS08}.  The tentative
detections of CS/C$^{34}$S $\sim$ 4 suggests opacities significantly
greater than unity in CS(2--1).

\noindent{\it CH$_{3}$OH \label{morph_ch3oh}---Methanol:} We observed
the low energy, blended set of $2_{1}-1_{1}$E, $2_{0}-1_{0}$E,
$2_{0}-1_{0}$A+ and $2_{-1}-1_{-1}$E thermal transitions of CH$_{3}$OH
(hereafter labeled the $2_{k}-1_{k}$ transition).  We cannot resolve
these transitions in this group, given the source linewidths.  
The $2_{k}-1_{k}$ transition of CH$_{3}$OH is very bright across the
nucleus, with peak temperatures reaching 0.28 K (Fig. \ref{FintiB}).
Of the species observed in this paper only HCO$^{+}$(1--0) is
brighter.  Remarkably though, the morphology of CH$_{3}$OH is 
completely different from what is seen in most of the other lines, including
$^{13}$CO(1--0) and HCN(1--0).  CH$_{3}$OH peaks strongly towards the
southern nuclear bar end (G and H) and the southern nuclear bar arm /
ring intersection region (F).  By contrast, emission is barely
detected near GMC D+E. The emission favors the right side of the
southern arm, which is the trailing, upstream side. At the bar ends
methanol abundances peak at $\sim 1.5 \times 10^{-8}$.  Towards
GMC D abundances drop by a factor of 5. The peak arm abundances
are a factor of two higher than those of IC 342 \citep[][]{MT05}. Galactic 
methanol abundances can reach these levels in small localized hot cores regions
\citep*[eg.][]{KS94,KDBWA97,BP97,T98,Num00,MB02}.  CH$_{3}$OH
velocities appear consistently blueshifted with respect to most of the
other transitions.  In the north the velocity shift is $\sim$ -10 km
s$^{-1}$, growing to $\sim$ -40 km s$^{-1}$ along the southern bar
end.  The shift towards the northern cloud is not significant, however 
the shift seen towards GMCs F and G appears real. This may result from an 
increased population in the higher excitation blueshifted lines that make up 
the quadruplet or somewhat different kinematics due to the presence of 
shocks (see section \ref{barshocks}).

\noindent{\it HNCO \label{morph_hnco}---Isocyanic Acid:} We observed
the K$_{-1}$ = 0 transition of the J = 5--4 rotational state
($5_{05}-4_{04}$) of this prolate, slightly asymmetric top. In the
higher resolution OVRO maps, HNCO peaks at GMCs F and G, with faint
emission along the bar connecting the two clouds (Fig. \ref{FintiO}).
When smoothed to match the resolution of the BIMA data, HNCO has a
very similar morphology to CH$_{3}$OH.  Over the 8$^{''}$ aperture used for
the spectra in Fig. \ref{Fspec}, HNCO has a peak brightness of 0.25
K, whereas over the 40 pc resolution achieved in the OVRO data peak
brightnesses reach 1.0 K.  No emission is detected towards GMC D 
in the high resolution map.  HNCO abundances reach
$X$$(\rm HNCO) \simeq 9\times 10^{-9}$ along the southern nuclear bar.
These abundances are typical of those observed on one to two orders of
magnitude smaller scales in Galactic massive dense cores
\citep[][]{ZHM00} and are even a factor of 3 larger than the HNCO
enhanced shock regions in IC 342 \citep[][]{MT05}.   
CH$_{3}$OH and HNCO in Maffei 2 have some of the most
pronounced morphological differences from CO yet seen in extragalactic
sources. These two species illustrate the importance of high spatial
resolution interferometric observations of chemical species.

\noindent{\it Other Transitions: \label{morph_others}} Table  
\ref{othermolT} presents the limits (2$\sigma$) for other potentially
detectable transitions within the observed bandwidth.  These include
the $^{13}$C substituted isotopologues of HNC(1--0) and
HC$_{3}$N(10--9) and two transitions of the radical HCO.  None of
these transitions are detected.  In the
case of the $^{13}$C isotopologues only the HN$^{13}$C(1--0) limits
are at all constraining (Table \ref{RatT}).  Towards the central ring  
and southern bar end 1$\sigma$ HNC(1--0)/HN$^{13}$C(1--0) line ratio
lower limits are $>$10--16, which in these locations are consistent
with or slightly above the corresponding CO isotopic line ratios
\citep[][]{MTH08}.  Therefore the HNC(1--0) opacities are lower than
$^{12}$CO(1--0) opacities.

To facilitate the exposition, we omit the transitions for each species
where possible in the remainder of the text. We discuss transition
dependent effects specifically where important. 

\subsection{Understanding the Molecular Maps: Principal Component 
Analysis \label{pca}}

The maps give a picture of the molecular cloud astrochemistry in the
nuclear region of Maffei 2. To interpret these maps, we need
to establish similarities and differences between the molecules.  The
correlations reveal trends in the chemistry, which in turn, can
suggest the physical driving forces responsible for these
characteristics and how they change across the nucleus.

As in MT05, we quantify morphological correlations amongst the images
via a principal component analysis (PCA).  PCA is a common technique
\citep*[eg.][]{MH87,K88,ED01,WJ03} used to reduce the dimensionality
of a dataset. The PCA simplifies the picture of molecular
distribution, reducing a large amount of information to a few images,
providing an excellent framework within which to study the complex
variations in molecular properties in Maffei~2. For a general
description of PCA applied to multi-transition molecular maps, see
\citet[][]{HS97} or for specific applications to extragalactic
molecular maps, see \citet[][]{MT05}. The PCA is dependent on the
particular set of data given to it; here, we are analyzing the
correlations of 3 mm lines of heavy molecules, so the results we obtain
pertain to the cool and relatively dense cloud population in Maffei~2.

To calculate the principal components for Maffei 2, the line maps were
remade to the same geometry and beamsize (that of HNC; Table
\ref{ObsT}), normalized and mean-centered.  Thirteen maps, including
$^{13}$CO(1-0), HCN(1-0) and 3 mm continuum from \citet[][]{MTH08},
plus the ten transitions of the nine species discussed in section 
\ref{morph_spec}, were sampled at 
1$^{''}$ intervals over the central $38^{''}\times 56^{''}$, making up
a 27664 element data-space.  The algorithm used to calculate the
principal components is that of \citet*[][]{MH87}.

The results of the PCA are tabulated in Tables \ref{pcacorT} and  
\ref{pcaT}.  Table \ref{pcacorT} shows the matrix of pair-wise  
correlation coefficients between all of the thirteen maps.  Table
\ref{pcaT} records the projection of each map onto the seven modeled  
PCs.  Fig. \ref{Fpcamaps} displays the maps of the first three (most
important) principal components (PCs).  These three PCs account for 80 
\% of the variation present in the data set.  The projections of each
map onto the first three PCs are shown vectorially in
Fig. \ref{Fpcavect}.

Inspection of Figure \ref{Fpcavect} reveals that PC 1 is most strongly
correlated with all maps. It represents a mix of the $^{13}$CO and HCN 
maps, implying that PC 1 is a good density-weighted column density map
of the low excitation molecules in Maffei~2.  This was also found in
IC 342; however, here $^{13}$CO projects along PC 1 more closely than
HCN, hence PC 1 is nearer to a pure column density map than its
analogue in IC 342.  All maps except the low S/N ratio C$^{34}$S show
significant correlation with PC1.

PC 2 characterizes the dominant variation seen once the distribution
of column density has been accounted for.  PC 2 distinguishes two major
distinct groups of molecules, those peaking towards the norther star forming region and
those peaking towards the southern nuclear bar end.  Species with
large positive projections on PC 2 (star formation peakers) include
HCO$^{+}$, HCN, HNC, C$_{2}$H and of course 3 mm continuum, which is
free-free emission.  Species with large negative projections (southern
bar peakers) include HNCO, CH$_{3}$OH and to a lesser extent
N$_{2}$H$^{+}$ and SiO.  PC 3 has a quadrupolar morphology dominated 
by SiO and C$^{34}$S.  PC 4 and beyond are at best marginally significant so 
we do not discuss them in detail.

The correlation matrix resulting from the PCA (Table \ref{pcacorT}) 
shows in more detail the correlations between various species.  The
following trends are noted.  

{\it (1) HCN, HNC, HCO$^{+}$ and 3 mm continuum are all extremely
tightly correlated, indicating a close association with massive star
formation.}  At the resolution of the PCA HCO$^{+}$ and HCN are
effectively perfectly correlated.  At three times higher spatial
resolution subtle differences become apparent   
(Fig. \ref{Fdenserat}). HCN is the map most tightly correlated with   
the 3 mm continuum.  Therefore, the (massive) star formation rate
versus HCN line luminosity relation seen in single-dish surveys
\citep[e.g.][]{GS04} persists to at least 100 pc scales in Maffei 2,
as noted in \citep[][]{MTH08}.  However, the correlation between
HC$_{3}$N(10--9) and 3 mm continuum is weaker than found for IC 342.

{\it (2) CH$_{3}$OH and HNCO are very well correlated with each other
and are strongly anti-correlated with the massive star formation.}
This leaves little doubt that on these GMC spatial scales the
chemistries of these two species are either directly coupled or are
mutually coupled to a specific chemical mechanism.  Moreover these
species show little correlation with 3 mm continuum and hence star
formation is not relevant to establishing the morphologies of these
species.  Nor are these species correlated with the standard dense gas
tracers, HCN, HCO$^{+}$ and HC$_{3}$N.  Fig. \ref{Fpv} displays the position-velocity 
diagrams of $^{13}$CO, C$_{2}$H and HNCO.  The strong differences between
the starburst tracers and CH$_{3}$OH or HNCO carry over to the kinematics.  
The 'parallelogram' originating from the central ring is nearly absent in 
CH$_{3}$OH and HNCO but remains pronounced in C$_{2}$H.

{\it (3) SiO and especially C$^{34}$S are less tightly correlated with
the other species.}  This may in part be due to their low S/N ratio
and that some of the structure that appears in these maps are
noise. However C$^{34}$S was also anomalous in IC~342 \citep{MT05}.

\section{Discussion} 

The pronounced variations in spatial morphology observed for the different
molecules across the central half kiloparsec of Maffei 2 are caused by
either differences in emissivity (excitation) or abundances
(chemistry).  In this section we discuss the possible influences on
the molecular emission that could lead to the observed variations.

\subsection{The Dense Gas Component: Diagnostic Ratios} 

The CO/HCN ratio is widely used to study the relative fraction of
dense gas in galaxies. \citet[][]{MTH08} used the CO/HCN ratio to
identify the sites of high dense gas fraction and concluded that the
$x_{1}-x_{2}$ orbital intersections and the eastern (star forming)
portion of the central ring are the locations of the high dense gas fraction.  The
fraction of dense gas drops towards the western ring and moving out
along the bar arms.

To study the nature of the dense component directly, line ratios
between dense gas tracers are required, since CO does not trace dense
gas.  Line ratios between bright, high critical density species, such
as HCN, HCO$^{+}$, HNC, HC$_{3}$N and CS are useful diagnostics of the
physical and chemical properties of the dense component of the ISM.
Much observational and theoretical effort has been dedicated to
understanding these diagnostic species in nearby starbursts
\citep[e.g.][]{HBM91,NJHTM92,SDR92,HB93,KMVOSOIK01,APHC02,GS04,INTOK07,
BHLBW08,GGPFU08,JNMSBKV09,BAMVM09,LACPMM11,AMMMB11}.  Unfortunately
interpreting these results can be difficult since most of these
studies have been done at low spatial resolution, either with
single-dish telescopes or in distant sources.  These ratios mix many
interstellar environments into one beam.

Now that we have access to an inventory of dense gas tracers at 30-100
pc resolution, we investigate changes in the HCN/HCO$^{+}$, HCN/HNC,
HCO$^{+}$/N$_{2}$H$^{+}$, HNC/HCO$^{+}$ and HCN/CS line ratios on GMC
scales across the nucleus of Maffei 2.  Fig. \ref{Fdenserat} displays
four of these line ratios.  Line ratios are tabulated in Table
\ref{RatT}.  Organized variations in all ratios are observed on these
scales and are discussed in terms of the CO/HCN ratio.

\subsubsection{HCN/HCO$^{+}$: Tracer of Intermediate Gas Densities\label{hcohcn}}

The HCN/HCO$^{+}$ ratio relates two of the brightest lines that trace
dense gas. It has become a popular diagnostic of gas 
with densities intermediate between CO and HCN  since 
HCO$^{+}$ has a critical density nearly an order of magnitude lower
than HCN  \cite[e.g.,][]{KMVOSOIK01,GGPC06,INTOK07,KNGMCGE08} .  
HCO$^{+}$ also has a formation route that maintains
significant abundances in relatively diffuse gas (see below), and it is an ion
making it more sensitive to collisions with electrons
\citep[e.g.][]{P07} (see Table \ref{MolP}).  As a result HCO$^{+}$ can
remain abundant and excited in diffuse gas \citep[][]{LL96}.
Theoretical modeling finds that HCN/HCO$^{+}$ is a strong function of
density for a given column density in the presence of ionizing
radiation, with ratios of unity for densities of $\sim$10$^{5}$
cm$^{-3}$ \citep[e.g.][]{MSI07,YWT07}.

There is, however, the complication of opacity. Despite its lower
critical density HCO$^{+}$ has a higher Einstein A$_{ij}$, implying
higher radiative efficiencies than HCN (and HNC).  For a
given abundance per line width, HCO$^{+}$ exhibits a line opacity 
nearly twice that of HCN and HNC.  For extragalactic studies it is
usually assumed that HCN abundances are significantly higher than
HCO$^{+}$. However, for the LTE abundances calculated in $\S$
\ref{abuintro} and \citet[][]{MTH08}, HCO$^{+}$ would have
approximately twice the observed abundance of HCN, contrary to what is
commonly assumed.  These abundances have large uncertainties but 
the two lines have similar energy levels.  The  
similarity in excitation requirements of the two lines makes it 
difficult to reverse this situation purely by changing gas
temperatures.  If the LTE abundances are correct then HCO$^{+}$ line
opacities must be at least as large as the presumably large HCN line 
opacities.  If opacities in both  lines are greater than unity
the diagnostic power of HCN/HCO$^{+}$ is limited.

Towards the nuclear bar in Maffei 2 the overall morphologies of
HCO$^{+}$ and HCN are similar. However, detailed differences in the
two transitions are revealed in their ratio.  HCN/HCO$^{+}$ ranges
from 0.6--2.2 over 30 pc scales in Maffei~2.  {\it These values cover
much of the range sampled in the single-dish surveys of many galaxies}
\citep[][]{KMVOSOIK01,GGPC06,KNGMCGE08}.  Both HCN \citep[][]{MTH08}
and HCO$^{+}$ fall off more rapidly with galactocentric distance than
does $^{13}$CO(1--0).  HCO$^{+}$ decreases more slowly than HCN in the
north (GMC B) but more rapidly along the southern bar arm.  HCN is
elevated relative to HCO$^{+}$ on the active eastern side of the central
ring, GMC G and a localized cloud just south of GMC F, while
HCO$^{+}$ is enhanced, relatively speaking, in the quiescent, western
ring region.  We detect clouds with CO/HCN vs. HCN/HCO$^{+}$ values
that reside in the "AGN dominated" portion of parameter space set by
global single-dish studies where clearly there is no AGN
\citep[][]{KMVOSOIK01,GGPC06}.  Similar statements apply for "star
formation dominated" parameters.  As a result interpretation of
single-dish values must be done with great care.

The enhancement of HCO$^{+}$ towards the western ring and HCN towards
the northeastern ring makes sense in the conventional
critical density context, since one expects the denser gas to be more
closely associated with the massive star formation.  However, the
elevated HCN/HCO$^{+}$ ratio towards the southern bar end is unusual.
A decreasing HCN/CO ratio suggests the fraction of dense gas in the
south is significantly lower, which should favor HCO$^{+}$.  That this
is not seen implies some other excitation mechanism or chemistry is in
play in the gas of the southwestern arm.  From the distribution of
shock tracers (section \ref{gasgrain}), GMC~G must be one of the 
clouds most effected by shocks.  It is tempting, therefore, to 
suggest that the southern bar end in Maffei~2 is an example of a
location of elevated HCN abundances in shocks.  In the Galaxy, bipolar
molecular outflows do show HCN much more strongly elevated than
HCO$^{+}$ \citep[e.g.][]{BP97}.  This scenario seems less favored when
the other dense gas tracers are incorporated.  Towards GMC G, both the
HCO$^{+}$/N$_{2}$H$^{+}$ and HCO$^{+}$/HNC ratios are anomalously low,
while the HCN/HNC ratio remains roughly constant relative to the rest
of the (non-star forming) nuclear disk.  This argues for a
depletion/destruction of HCO$^{+}$ rather than an enhancement of HCN.
The decrease in HCO$^{+}$ intensity here may stem from the shock in a
more indirect way.  The passage of a shock alters the chemical state
of the molecular gas, including a dramatic increase in gas-phase
water, which is effective at destroying HCO$^{+}$ \citep[][]{BMN98}.
The main difficulty with this mechanism is that H$_{2}$O is also
expected to destroy N$_{2}$H$^{+}$, which is not observed.

An alternate and perhaps simpler explanation for the elevated
HCN/HCO$^{+}$ ratio and the brightness of high density tracers like
HNCO and N$_{2}$H$^{+}$ at the southern bar end is that while the
dense gas fraction decreases along the southern bar arm, the density
of the dense component does not.  In this context the correlation
between the dense gas tracers and CO/HCN towards the star forming ring
would imply that the dense gas fraction and dense gas density are
coupled there.  The difference between the two environments could be a 
result of the different dynamical environments at each location.
Towards the southern bar end one can envision a situation where the
shocks (section \ref{barshocks}) produce a low filling factor, high
density component that explains the behavior of both ratios.  The fact
that the HNCO/CH$_{3}$OH ratio remains high
gives additional direct evidence that densities of the shocked gas are
quite high towards GMC G (section \ref{relshocks}).

In summary, while the CO/HCN ratio suggests a rapidly decreasing dense
gas fraction along the arms \citep[][]{MTH08}, The HCN/HCO$^{+}$ ratio 
suggests that the actual density does not fall significantly.  Clearly not all the properties
of the dense component are traced in the CO/HCN ratio.  This further
emphasizes the fact that the {\it CO/HCN measures the fraction of the gas
that is dense, not necessarily the actual density of the gas.}  Ratios
that sample only dense gas, such as HCN/HCO$^{+}$, are needed to 
determine the physical conditions of the dense component.

\subsubsection{HCN/HNC: Proposed Tracer of Warm Gas \label{hcnhnc}}

The HCN/HNC line ratio is another dense gas ratio of increasing
popularity, in which high values indicate hot gas and/or PDR conditions.  
Essentially identical A$_{ij}$, collision coefficients,
upper energy states and frequencies make the ratio particularly
insensitive to changing excitation.  Dissociative
recombination of HCNH$^{+}$ dominates the formation of both HCN and
HNC in cool, dense clouds, producing both species in equal amounts
\citep[e.g.][]{HYMO98}.  So in these environments the abundances of
the two species are locked together.  Additionally both transitions
likely have significant opacities.  It is therefore not surprising
that this ratio remains near unity over a wide range of conditions.
When gas temperatures get large HNC can be converted to HCN via HNC
$+$ H $\rightarrow$ HCN $+$ H, while in diffuse gas / PDR-like
environments, the reaction N +CH$_{2}$ $\rightarrow$ HCN $+$ N becomes
a major additional HCN source \citep[e.g.][]{SWPRFG92,TPM97}.  

Fig. \ref{Fdenserat} (Table \ref{RatT}) displays the HCN/HNC intensity
ratio map.  Indeed the ratio remains fairly constant across the
nucleus with values of $\sim$1.1 $\pm$0.3.  Variations in the ratio
are only at the 30 \% level over most of the nucleus, with the highest
values on the star forming eastern ring and south of
GMC~F, which suggests mild enhancement by PDRs. Ratios significantly
below unity on these small spatial scales are not seen anywhere across
the nucleus. Hence there is no evidence for anomalous enhancement of
HNC relative to HCN that has been found in some ULIRGs
\citep[][]{APHC02}. In IC 342, the HCN/HNC intensity ratio is also
fairly constant, with weak enhancement seen towards the PDR GMC
\citep[][]{MT05}.

In summary, at least in Maffei 2 and its less active cousin, IC~342, the HCN/HNC
line ratio has limited diagnostic power.  Its diagnostic power becomes
stronger for molecular gas with extreme temperatures and density, such
as LIRGs and ULIRGs, where the chemistry is driven away from
dissociative recombination of HCNH$^{+}$.

\subsubsection{HCO$^{+}$/N$_{2}$H$^{+}$: Tracing the Dense Gas \label{hcon2h}}

Both HCN and HCO$^{+}$ (and to a lesser degree HC$_{3}$N) are observed
to maintain large abundances in PDRs, shocks and other energetic
regions.  By contrast, in the Galaxy N$_{2}$H$^{+}$ is observed to
avoid PDR regions, maintaining high abundances only in dense,
quiescent, dark cloud-like conditions \citep[][]{WZW92}.  In dense,
quiescent gas both N$_{2}$H$^{+}$ and HCO$^{+}$ form from proton
transfer reactions of H$_{3}^{+}$ with N$_{2}$ and CO, respectively:

\begin{equation} 
\label{En2h}
H_{3}^{+}  ~+~ N_{2} ~\rightarrow ~ N_{2}H^{+} ~+~ H_{2},
\end{equation}
\begin{equation} 
\label{Ehco}
H_{3}^{+}  ~+~ CO ~\rightarrow ~ HCO^{+} ~+~ H_{2}.
\end{equation}
Given the reasonable assumption that the rate coefficients for these
two reactions, $k_{\ref{En2h}}$ and $k_{\ref{Ehco}}$, are
approximately equal \citep[][]{WAMM07}, then
$X$(N$_{2}$H$_{+}$)/$X$(HCO$^{+}$) $\simeq$ $X$(N$_{2}$)/$X$(CO).
However, only HCO$^{+}$ maintains a relatively rapid formation route
in diffuse or PDR gas (C$^{+}$ + OH $\rightarrow$ CO$^{+}$ + H
followed by H$_{2}$ + CO$^{+}$ $\rightarrow$ HCO$^{+}$ + H).  No
corresponding route exists for N$_{2}$H$^{+}$ in PDRs
\citep[][]{T95b}. This explains the observed behavior of the
HCO$^{+}$/N$_{2}$H$^{+}$ line ratio (Fig. \ref{Fdenserat} and Table
\ref{RatT}).  Towards the PDR regions associated with the active star formation
sites, HCO$^{+}$ can maintain appreciable abundances while
N$_{2}$H$^{+}$ is readily destroyed.  The fact that this ratio, which
reflects dense gas properties, is modified by the starburst PDRs
implies that towards this region (GMC E and the eastern ring) PDRs
penetrate the dense gas.  {\it The strong correlation between high
HCO$^{+}$/N$_{2}$H$^{+}$ ratios and PDR tracers like C$_{2}$H,
suggests that this ratio may be a powerful new probe of {\bf dense} PDRs in
external galaxies.}

Along the arms, $X$(N$_{2}$H$^{+}$)/$X$(HCO$^{+}$)
abundances are typically $\sim$0.1-0.15, with values reaching up to 
$\sim$0.33 towards the southern bar end.  If a standard CO abundance of $X$(CO)
$\simeq$ 10$^{-4}$ is adopted, $X$(N$_{2}$) abundances of $1 - 3
\times 10^{-5}$ are implied.  If metallicities are solar
\citep[e.g.][]{AG89} and all N is in the form of N$_{2}$, then
$X$(N$_{2}$) $\simeq ~4\times 10^{-5}$.  Evidently, in the more
quiescent dense clouds of the nucleus of Maffei 2, at least a quarter of
all nitrogen is in molecular form.  Using the above $X$(N$_{2}$)
estimation together with Table \ref{AbuT} gives
$X$(N$_{2}$)/$X$(N$_{2}$H$^{+}$) $\simeq ~ 2\times 10^{4}$ for these
regions.

Following the analysis of \citet[][]{MT05},
$X$(N$_{2}$)/$X$(N$_{2}$H$^{+}$) provides indirect constraints on the
cosmic ray ionization rate, $\zeta$.  Adopting
$X$(N$_{2}$)/$X$(N$_{2}$H$^{+}$) $\sim ~ 2\times 10^{4}$ and the
extremely conservative lower limit, $X(e^{-})$ = $X(ion)$ $\ge ~
X($N$_{2}$H$^{+}$) ~+ ~ X(HCO$^{+}$) $\ge 6-9 \times 10^{-9}$, Figure
9 of \citet[][]{MT05} indicates $\zeta /n_{H_{2}} > 10^{-21}$ s$^{-1}$ cm$^{3}$.  Less
conservatively, for $X(e^{-}) \simeq 10^{-7}$ typical of Galactic star
forming regions \citep[e.g.][]{CWTH98}, $\zeta /n_{H_{2}} >
10^{-20.5}$ s$^{-1}$ cm$^{3}$.  Therefore, $\zeta$ is at least the Galactic value if the
density of the gas that the N$_{2}$H$^{+}$ and HCO$^{+}$ emission
originates in is $n_{H_{2}} ~>~ 10^{4}$ cm$^{-3}$.  These constraints
are comparable to though somewhat stronger than observed for IC 342.

\subsubsection{The CS(2--1)/HCN(1--0) and CS(2--1)/C$^{34}$S(2--1) Ratios \label{cs_hcn}}

\citet[][]{KNSS08} has published a map of CS(2--1) towards the nucleus
of Maffei 2 with similar spatial resolution to the data presented
here.  We make use of this map to compute the CS/C$^{34}$S and CS/HCN
ratios towards the GMCs.  The CS/HCN intensity ratio is fairly
constant across the nucleus except towards the southern arm (GMC G),
where it is a factor of two lower.  If strong chemical evolution has
not altered the $^{32}$S/$^{34}$S ratio, the C$^{34}$S emission
implies the regions of high CS opacity are along the southern arm and
not at the CS peak near the massive star formation.  This is unexpected. For a
$^{32}$S/$^{34}$S ratio of 23, LTE line ratios imply CS opacities of
$\sim$6 along the southern arm.  Comparing the observed peak
temperature of C$^{34}$S gas to optically thin emission with T$_{ex}$
= 10 K yields a filling factor, $f(C^{34}S)\simeq 0.018$.  Given the
differences in beam size this filling factor is intermediate between
$^{13}$CO and HCN \citep[][]{MTH08}.  Towards the GMCs D+E the
larger CS/C$^{34}$S ratio implies that CS opacities here must be less
than 1.4.  Since the intensity of the main isotopologue peaks towards
this region, gas temperatures or filling factors are much
larger here.  So in order to explain the morphology of both CS and
C$^{34}$S, it is required that CS be warmer and more diffuse towards
GMCs D+E when compared to the GMC G.  In summary CS
appears compatible with the interpretation based on HCN/HCO$^{+}$ 
that the southern bar end has high density but low dense gas fraction.

\subsubsection{HC$_{3}$N(10--9)}

HC$_{3}$N has a critical density that is comparable or slightly higher
than HCN but very likely has low opacity due to its lower abundance.
Moreover its levels are more closely spaced and so population is distributed
over a wider range of levels, making any one level more sensitive to
changes in excitation.  HC$_{3}$N originates primarily from the
arm-ring intersection regions similar to HCN and HCO$^{+}$, although
perhaps peaks slightly farther from the center of the galaxy than does
HCN.  Using the kinematics to further resolve the emission suggests
that HC$_{3}$N is tracing the inner portion of the arm rather than the
$x_{1}-x_{2}$ intersection region.  Evidently, densities increase
along the arms as the central ring is approached.  Unlike HCN,
HC$_{3}$N weakens just before the strong star formation is reached at GMC D
+ E and just north of GMC F.  Potentially this results from warmer
temperatures depopulating the J = 10 level towards the starburst.

\subsection{PDR Chemistry and the Nuclear Starburst \label{pdr}}

Ionizing radiation from massive stars impinging on nearby molecular
clouds changes their chemical composition.  Large abundances of
ionized carbon in PDRs drive a rich hydrocarbon chemistry on cloud
surfaces.  One of the simplest species from such regions is C$_{2}$H.
C$_{2}$H is formed from rapid dissociative recombination of
C$_{2}$H$_{2}^{+}$ coming from reactions between C$^{+}$ + CH$_{3}$
\citep[][]{WBGLS80,SD95}.  {\it C$_{2}$H chemistry is particularly
closely tied to ionized carbon and is an excellent probe of these
PDRs.}  Its usefulness as an extragalactic PDR tracer was demonstrated
in \citet[][]{MT05}.

After accounting for opacity effects (see $\S$\ref{morph_spec} for 
C$_{2}$H(1--0; $\frac{1}{2}-\frac{1}{2}$) for an optically thin C$_{2}$H transition)
C$_{2}$H pervades much of the molecular clouds in the northern star forming
region.  Lower level C$_{2}$H emission extends over the rest of the
nucleus but avoids the cloud cores, instead favoring the side of the
dense cloud facing star formation (traced here by 2 cm radio
continuum) (see Fig. \ref{cchoutf}).  This is particularly clearly
seen west of GMC F and northeast of GMC G.

As discussed in $\S$\ref{hcon2h}, the dramatic increase in the
HCO$^{+}$/N$_{2}$H$^{+}$ line ratio across the central ring shows not
only that PDRs are prevalent near the young, star forming region but that the PDRs
penetrate much of the dense gas there. Given the distinct region of
altered chemistry, constraints can be placed on the range of
influence of the current burst of star formation.  For the northern star forming region, the
region of elevated HCO$^{+}$/N$_{2}$H$^{+}$ ratio is 80 pc diameter in
extent. We assume that the volume of the region emitting strong PDR
emission reflects the volume of molecular gas engulfed within
HII regions that have expanded into the ambient medium due
to over-pressure. This should be reasonable on GMC scales.  For an
ionized gas sound speed of 10 km s$^{-1}$, N$_{Lyc} = 5\times 10^{51}$
s$^{-1}$ and a uniform density of $n_{H_{2}}\simeq 10^{2.75}$
cm$^{-3}$ typical of the central ring \citep[][]{MTH08}, a standard
expanding HII region would reach a radius of 40 pc in $\sim$7 Myr.
While this analysis is sensitive to the exact density distribution, it
is interesting that this timescale for dynamical evolution of GMCs due
to pressure is an order of magnitude longer than characteristic
chemical timescales and consistent with the timescale of massive star
evolution.

\subsection{The Molecular Outflow along the Minor Axis in PDR Gas \label{pdrout}} 

{\it C$_{2}$H emission appears to trace a molecular outflow along the
minor axis (Fig. \ref{cchoutf}).}  The northwest extension of this
bipolar flow is redshifted by $\sim$ 15 km s$^{-1}$ and the southeast
blob blueshifted by $\sim -$15 km s$^{-1}$ with respect to the velocity 
of the corresponding in-plane material.  These velocity shifts have
the correct sign for outflow given Maffei 2's geometry (northern arm
being nearer to us, and the arms trailing).  There is tentative
evidence for the flow in HNC and SiO \citep[and possibly
CO(1--0);][]{MTH08} but the molecular outflow is not obvious in any of
the other species.  This is understandable given the strength of
C$_{2}$H in the outflow reflects its relatively low critical density
and its ability to maintain high abundances in the presence of
ionizing radiation.

The molecular outflow in Maffei 2 does not appear to follow the distribution 
of ionized gas traced by 2 cm.  This is different from what is seen in the 
much more pronounced outflow in M82 \citep[][]{WWS02}. The 2 cm radio 
continuum emission appears to trace the walls of Maffei 2's outflow
\citep[Fig. \ref{cchoutf};][]{TH94}.  The C$_{2}$H emission must trace
entrained molecular material being driven out directly.  The
explanation for the 2 cm emission along the edges remains unclear but
may be due to a path length effect of a cone of ionized gas
surrounding the molecular flow.  Since evidence is seen for a small
minor axis outflow it is possible to obtain a second estimate of the
characteristic age by assuming that the gas traced in C$_{2}$H is free
to flow ballistically along the minor axis.  The extent of the outflow
is $\pm$200 pc.  For an inclination angle of 67$^{o}$ and the measured
line of sight velocity, the time of flight corresponds to 6 Myr.  This
is a very young outflow.  The agreement with the characteristic
in-plane timescale is excellent, suggesting that the current starburst
may have an age of $\tau \sim$ 6 - 7 Myr. 
 
\subsection{Shocks and Gas-Grain Chemistry: CH$_{3}$OH, HNCO and SiO \label{gasgrain}}

Elevated gas-phase abundances of species whose formation is sensitive
to the grain processing, such as CH$_{3}$OH, HNCO, or 
SiO, are important probes of hot core gas and
shocks.  As noted in \citet[][]{MT05}, and illustrated even more clearly here, these
species do not trace hot cores.  Typical Galactic hot cores, even
significant collections of them cannot produce the required emission
necessary to explain the intensities on these large scales.  
Moreover, we detect no spatial connection between CH$_{3}$OH and 
HNCO, and star formation, if anything, we see an anticorrelation.  We can, therefore, 
safely conclude that the above three species are not tracing hot core 
emission or outflows from  young stellar objects but must originate in large scale 
shocks.  Inter-comparisons can be used to identify the responsible shock 
source \citep[][]{MT05,RMRMRD06,RCMWA06,MRMM08} and place constraints on 
shock properties such as shock strength, type of shock and liberated grain /
mantle composition \citep[][]{UGMFN06}.

In this dataset three shock tracers, CH$_{3}$OH, HNCO and SiO, are
mapped.  Each is sensitive to somewhat different shock properties.  It
is expected that elevated gas-phase SiO results from release of Si
from destruction of silicate grain cores \citep[e.g.][]{MBF92}.  That
the grain core must be disrupted requires 'strong' shocks capable
of sputtering grain cores.  High observed abundances of
gas-phase CH$_{3}$OH likely come from direct liberation of CH$_{3}$OH
that has formed on grain surfaces by complete hydrogenation of CO
\citep[e.g.][]{MHC91,BLWC95}.  Liberation of mantles does not require 
shocks as strong as those that effect the grain core.  Thus the relative 
enhancement of SiO versus CH$_{3}$OH will depend on the strength of 
the shock. Grain core destruction requires shock
velocities of $\gsim$25 km s$^{-1}$ whereas shock velocities of only 
$\gsim$ 10-15 km s$^{-1}$ are required for
significantly elevated gas-phase CH$_{3}$OH abundances
\citep[e.g.][]{CHH97,BMN98}.  It is reasonably well established that
HNCO is a mantle species along with CH$_{3}$OH
\citep[][]{MT05,RTGB10}.  But its formation mechanism remains unclear.
The most commonly proposed mechanism is partial hydrogenation of OCN
on grain surfaces followed by direct liberation \citep[e.g.][]{HH93}.
Other suggested mechanisms include acid-base reactions between CO and
NH$_{3}$ and UV photolysis \citep[][]{SG97,KTBSW01}.  Thus if HNCO
forms on grain mantles and is liberated directly then its unsaturated
nature implies that it originates in grain mantles with different
chemical composition from CH$_{3}$OH (however see \citet[][]{TFMM10}
for a possible formation mechanism in more saturated
ices). Unsaturated ices are favored when gas accretes in atomic hydrogen-poor
environment whereas saturated ices favor atomic hydrogen-rich accretion
\citep[e.g.][]{CTR97,WK02}.  Hydrogen-poor conditions reflect accretion at
high densities when almost all hydrogen is in molecular form.  If it
is assumed that the shock simply injects molecules into the gas phase,
the HNCO/CH$_{3}$OH abundance ratio is then sensitive to gas physical
conditions during ice mantle formation.

Chemistry is not the only effect that influences the relative
abundance of these three species.  Excitation is also important.
The excitation requirements of SiO(2--1), CH$_{3}$OH($2_{k}-1_{k}$)
and HNCO($5_{05}-4_{04}$) transitions differ (see Table
\ref{MolP}). The critical density of the HNCO transition is an order
of magnitude higher than SiO and CH$_{3}$OH.  Furthermore the
energy of the upper state in HNCO($5_{05}-4_{04}$) 
is more than twice that of SiO and
higher than two of the three main transitions that make up the 
CH$_{3}$OH line.  Therefore, CH$_{3}$OH and SiO lines will be favored
over HNCO in lower density gas.  Also in the presence of ionizing
radiation each species dissociates at different rates.  HNCO is
photodissociated at a rate twice that of CH$_{3}$OH and $\sim$30 times
that of SiO \citep[][]{RJLD91, SD95}.

Thus considering the distribution of CH$_{3}$OH, HNCO, and SiO
can give insights into both the chemical and physical conditions
present in the shocked gas and potentially tell us about nature
of the shocks themselves.  Below we discuss the overall structure of
the shock tracers in the context of nuclear bar models, then we
analyze the subtle differences between each shock tracer to further
constrain shock properties.

\subsubsection{Overall Shock Tracer Morphology and Bar Structure \label{barshocks}}

In Maffei 2 the outer bar arms dominate the overall distribution of
the shock tracers CH$_{3}$OH, HNCO, and SiO, whereas emission from the
star forming ring is minimal.  This is very different from the morphology
of CO, HCN and the PDR tracers. The bar ends are especially bright in
the shock tracers.  (The faintness of shock tracers along the northern
arm is presumably due to the fact that there is much less molecular
gas there, particularly at what would be the location of the northern
bar end.)  It is perhaps surprising that it is the bar ends and not
the inner arms that have the largest fraction of the gas influenced by
shocks.  This provides constraints on the nature of gas flows in
Maffei 2.

Different models exist for describing the structure of gas bars, each
predicting different shock strengths and locations.  The models fall
into two broad classes, orbit-crowding models and hydrodynamical
flows.  Orbit-crowding models treat the gas as distinct clouds
orbiting on essentially ballistic orbits set up by the potential
\citep[e.g.][]{CG85, BRSBC94}.  The arm morphology of the bar is then
determined by where orbits converge.  In hydrodynamical models the gas
is treated as a fluid that responds to the stellar potential and is
susceptible to hydrodynamical forces such as pressure and viscosity.
In such models, gas flows along streamlines with shocks occurring
where the streamlines collide \citep[e.g.][]{A92,PST95}.  Both models
predict similar morphological distributions but rather distinct shock
properties \citep[e.g.][]{RSV99}.  In orbit-crowding models the bar
arms result from molecular clouds moving along converging orbits. In
such models only weak shocks are expected because the bulk of the
velocities are parallel to both clouds (with the possible exception of
arm / nuclear ring intersections).  Moreover such models do not
isolate the bar ends as locations of strongly convergent orbits and so
it is not expected that these location would be sites of vigorous
shock activity.

Hydrodynamical models, on the other hand, show gas flow streamlines
that change directions dramatically.  The places where flows converge
are accompanied by strong changes in velocity, and hence strong
shocks.  In hydro models, gas drifts slowly inward along bar-aligned
$x_{1}$ stellar orbits (or in the case of Maffei 2, from radial inflow
along the outer bar) until the gas reaches the innermost, cuspy
$x_{1}$ orbit (of the nuclear bar).  The cusps of these orbits lie at
the bar ends.  There the gas is compressed, shocks and plunges nearly
radially inward along the bar arm until it reaches the much further
in, perpendicular to the bar $x_{2}$ orbitals, where it is again
shocked, loses angular momentum and transitions onto the $x_{2}$
orbitals.  Some high angular momentum gas 'sprays' off the inner
$x_{2}$ orbits, swinging out and colliding with the arm gas flow
\citep[e.g.][]{A92,PST95,KSJFMW11}.  The amount, location and relative
velocity of gas in the spray region depends on bar parameters.

{\it In summary, the strength of the emission from the shock tracers
CH$_{3}$OH, HNCO, and SiO along the bright bar ends favors the
hydrodynamical bar models, as orbit-crowding models do not predict
strong cloud collisions at the bar end whereas they are sites of
shocks in hydrodynamical models.}

\subsubsection{Relative Shock Tracer Morphologies and Bar Properties \label{relshocks}}

What about changes in shock properties across the nuclear bar?
Fig. \ref{hncoch3oh} displays the HNCO/CH$_{3}$OH, SiO/CH$_{3}$OH and
SiO/HNCO peak temperature ratios smoothed to 8$^{''}$ and sampled at
the GMCs.  Given the significant difference in the native resolution
of the three datasets and the fact that they were obtained with two
different interferometers, only the clearest trends are discussed.

As demonstrated by the  principal component analysis, 
the morphologies of the three species are 
strongly correlated, implying that we have a robust
measure of the locations of shocked gas.  {\it So similar are the maps
that only very subtle differences between HNCO, CH$_{3}$OH and SiO
appear, suggesting that shock conditions are uniform across the
nucleus.}  With the possible exception of the uncertain SiO ratios
towards GMC B, the arm clouds (C, D, F, G) have effectively constant
HNCO/CH$_{3}$OH ($\sim 1.1\pm 0.2$), SiO/CH$_{3}$OH ($\sim
0.20\pm0.02$) and SiO/HNCO ($\sim 0.19\pm0.03$) ratios.  Only towards
GMC E and the off arm spray regions (GMC H and just south of the
western ring) are different ratios observed.  Towards GMC E (the
northern star forming region), HNCO/CH$_{3}$OH is lower and the SiO ratios are higher
relative to the rest of the arm.  In the off arm regions
HNCO/CH$_{3}$OH ratios are depressed while the SiO/CH$_{3}$OH ratio is
elevated and SiO/HNCO are elevated (spray region) or flat (GMC H).

From these ratios, three main points are concluded.  1) While GMC G
(the bar end) is the brightest source in the nuclear bar, indicating
that the largest fraction of gas along this line of sight is shocked,
evidently shock strengths are not enhanced here.  SiO to mantle
species have the same basic ratio at the bar end as it has over the
rest of the bar.

2) Towards GMC E, the decreased intensity of HNCO
relative to CH$_{3}$OH and SiO is consistent with what is expected
from photodissociation.  From section \ref{pdr}, we know that the
dense gas here is permeated by UV photons for the young star
formation.  The shock tracer ratios further confirm this.  Because of
this, the absence (or faintness) of shock tracers towards the central
ring and in particular the northern $x_{1}-x_{2}$ orbit intersection
cannot be taken as evidence of absence of shocks here.

3) The spray region upstream (counterclockwise) from the main bar arms
favor CH$_{3}$OH and to a lesser degree SiO relative to HNCO.  
It is possible that because gas is more diffuse in this region grain
mantles are H-richer than in the denser arm. However, we favor a simpler
explanation associated with changing excitation. The spray regions
have lower density hence the higher critical density HNCO transition
is expected to be weaker relative to the lower critical density,
CH$_{3}$OH and SiO transitions.  That the $5_{05}-4_{04}$ transition
of HNCO preferentially favors the dense clouds relative to more
extended CH$_{3}$OH explains why larger beam single-dish multi-line
observations derive an excitation temperature for HNCO that is 
nearly a factor of $\sim 2$ larger than derived for the CH$_{3}$OH
transitions in Maffei 2 \citep[][]{HMH97,MMM09}.  Note that the
excitation temperature of both species is quite low (T$_{ex}$ $< $15
K), implying that the shock regions remain cool or the densities are
low enough for these transitions to be sub-thermal.

{\it In summary, we conclude that there is little evidence for
strongly varying grain chemistry across the shocked regions on
8$^{''}$ scales.  The small level of variation seen is consistent with
changing physical conditions (radiation field and gas density).}

\subsection{The Chemistry of Spiral Nuclei: Maffei 2 vs. IC 342 \label{overall}}

The overall nuclear structures of Maffei 2 and IC 342 are similar.  
Both have extended bar-like arms that terminate on a central ring that is
actively forming stars.  The molecular gas morphology of IC 342 is
consistent with that of a gaseous bar \citep{TH92}.  However, because
of the very small central ring and ambiguous kinematics, other
conclusions including unbarred spiral arms coupled with massive star 
feedback remain feasible \citep[][]{SBMC08}.  On the other hand Maffei
2 is clearly a nuclear bar based on both its morphology and kinematics
\citep[e.g.][]{MTH08,KNSS08}.  Given the potential dynamical
similarities of the two nuclei a comparison of the chemistry of the
two can provide important insights on the generalizability of the the
overall chemical picture developed in \citet[][]{MT05}.

The statistically robust principal component analysis demonstrates that the 
chemistry of the two nuclei are remarkably consistent. Maffei 2 can be defined
chemically as a highly inclined copy of IC~342 (Fig. \ref{Fgenchem}).
In both systems the first principal component characterizes the 
density-weighted column density, and peaks along the inner arms and 
central ring.  The second principal component separates the main 
chemical forces responsible for controlling
observed abundances, namely shocks versus PDRs.  PDR species
dominate close to the young star forming region while the shock tracers peak along the
arm, with preference for the cuspy bar ends.  As a result, the bar
driven star formation / chemical scenario outlined in \citet[][]{MT05}
must apply also to Maffei 2's nuclear bar.

However the chemical picture is not precisely identical, and these differences 
give added insight into the relative differences in evolutionary phase of the two 
nuclei.  The main difference between the two galaxies 
is in the distribution of N$_{2}$H$^{+}$.  
In IC~342 the PDR tracers are confined, appearing to be primarily cloud 
surface features, lining the inner ring.  Significant N$_{2}$H$^{+}$ (and
HNCO) seen near the young star forming region in IC~342 implies that its main 
star forming region influences the dense gas only locally, on scales of 50 pc or
less.  {\it Whereas in Maffei 2, the N$_{2}$H$^{+}$/HCO$^{+}$ ratio 
and photodissociation of HNCO clearly demonstrate that the star formation
has begun to dramatically alter the chemical state of the dense
molecular clouds in its immediate environment, not just the surfaces.}

Evidently the central molecular ring in Maffei 2 is structurally
different (lower clump density) and that towards GMC E star formation
is somewhat more evolved than the main burst in IC 342.  This is
consistent with the steeper radio spectrum towards GMC E
\citep[][]{TH94,TTBCHM06}.  The crude estimates of timescale required to
alter the chemistry of the dense component is $\sim$ 7 Myrs.  This is
a reasonable timescale compared to stellar evolution of the burst in
Maffei 2 and the kinematics of the outflow.  However, this timescale
is still very short; short enough that even this 'evolved' part of the
burst is in its early stages. The star formation extending to the
north and south of GMC E is likely as young as the embedded star
formation in IC 342. 

In IC~342 the peak temperature ratios between SiO, CH$_{3}$OH and
HNCO, are similar to those seen  in Maffei~2. The PDR GMC in IC~342
has slightly lower HNCO/CH$_{3}$OH and elevated SiO/HNCO ratios.  
This is exactly the behavior exhibited by the PDR-modified GMC E in Maffei 2, lending
strong support for the influence of photodissociation on shock tracer
brightness.  The shock chemistry dominated cloud in IC 342 has
slightly elevated HNCO relative to the other species, making its set
of ratios a match within the uncertainties of GMC G. This agreement
suggests that both the grain composition and shock properties are the
same for these structurally similar locales.  However, since
brightnesses are sensitive to radiative transfer effects, one must be
careful over-interpreting ratios if gas excitation is changing
rapidly.  Maffei 2 and IC 342 do have rather similar CO excitation
conditions \citep[][]{MTH00,MT01,MTH08}.

In summary, we find that when the dynamical environments are
similar, as they are in IC~342 and Maffei~2, 
a consistent chemical picture begins to emerge.  
Gas chemistry is confirmed as a powerful method for
isolating the physical mechanisms that influence gas conditions in
spiral nuclei.  In future papers in this series, we will broaden the
range of galactic environments studied to assess how far chemistry can
be pushed as a diagnostic of sub-beam physics.

\section{Conclusions \label{conc}}

This paper presents 2$\arcsec$ - 10$\arcsec$ imaging of  transitions 
from nine separate molecules across the nuclear bar in Maffei 2 obtained with
the BIMA and OVRO millimeter interferometers.  The results are
compared with the first paper in the series on IC 342 to test the
generalizability of the chemistry discovered there and to constrain
differences in structure between the two nuclei.  Our main conclusions
are as follows:

\begin{enumerate}
\item Morphologies of the observed transitions exhibit both dramatic
(e.g., C$_{2}$H vs. CH$_{3}$OH) and subtle (e.g., HCN vs. HCO$^{+}$)
spatial differences indicative of the varying dominance of shocks vs. 
radiation fields as drivers of the chemistry in individual clouds. 

\item A principal component analysis reveals that HCN,
HNC, HCO$^{+}$ and 3 mm continuum are all extremely tightly correlated,
indicating a direct connection to massive star formation.  

\item CH$_{3}$OH and HNCO are tightly correlated with each other
and are strongly anti-correlated with the tracers of massive star formation; 
based on their spatial distribution, it appears that these two molecules 
are formed by liberation of grain mantles in the presence of bar shocks.

\item Variations in the ratios amongst the popular $\lambda\sim$ 3 mm dense gas tracers, 
HCN, HCO$^{+}$, HNC, CS and HC$_{3}$N, are at or less than a factor of three, 
unlike the order of magnitude variations exhibited by the chemistry-sensitive
molecules.  This indicates that variations in excitation in the cool,
dense gas are much less dramatic than variations in dense gas chemical
conditions.  

\item Localized variations in the ratios HCN/HCO$^{+}$, HCN/HNC, HCN/CS and
HCN/CO can be explained by variations in density. The dense gas ratios indicate 
that the highest density gas is found at the $x_{1}-x_{2}$ orbit 
intersections, the lowest density gas is along the non star-forming 
western portion of the ring, and intermediate density gas is found  
near the star formation. Although high density gas is found throughout 
the full extent of the nuclear bar arms, the dense gas fraction falls at larger 
radii from the nucleus. 

\item There is no evidence for X-ray dissociation regions or AGN influences on the 
gas in the center of Maffei~2, specifically, as traced by the HCN/HCO$^{+}$ ratio. 

\item The strong correlation between high HCO$^{+}$/N$_{2}$H$^{+}$ ratios
and the PDR tracer molecule, C$_{2}$H, suggests that this ratio may be a powerful 
new probe of dense PDRs in external galaxies. Given the high critical
density of both HCO$^{+}$ and N$_{2}$H$^{+}$ in comparison to
C$_{2}$H, the elevated ratio demonstrates that PDR gas towards the 
strong star forming regions permeate the dense clumps.  When combined with the
morphology of C$_{2}$H this ratio may provide useful insights into the
relative density of PDR gas.

\item C$_{2}$H emission reveals a molecular outflow along the minor axis
that has entrained molecular gas.  Wrapped around the outside of this
outflow is ionized gas traced in 2 cm radio continuum.  Assuming
ballistic motion, the morphology and kinematics are consistent with a
very young age of 6-7 Myrs for the outflow.

\item The bar induced orbital shocks in Maffei 2 are traced in three
species, HNCO, CH$_{3}$OH and SiO.  All three species support the
conclusion that the nuclear bar end is the site of the largest
fraction of shocked gas.  That this location exhibits such strong
shock properties argues in favor of hydrodynamical bar models.
\end{enumerate}

\acknowledgements 
 
We appreciate the support and assistance of the faculty, staff and
postdocs at BIMA and OVRO during the observations.  We L. E. Snyder
and E. C. Sutton for helpful discussions.  The referee is acknowledged 
for a thorough and very helpful report.  DSM acknowledges support
from NSF AST-0506669, AST-1009620 and the National Radio Astronomy
Observatory which is operated by Associated Universities, Inc., under
cooperative agreement with the National Science Foundation.
Additional support for this work is provided by NSF grant AST-0506469
to JLT.

\clearpage
 
 \begin{deluxetable}{lccccccc} 
\tabletypesize{\scriptsize}
\tablenum{1} 
\tablewidth{0pt} 
\tablecaption{Molecule and Column Density Parameters\label{MolP}} 
\tablehead{ 
\colhead{Molecular}  
&\colhead{$\mu$}
&\colhead{A/B/C}
&\colhead{$S_{ul}g_{K}g_{I}$ }
&\colhead{$E_{u}$}    
&\colhead{$^{H_{2}}n_{cr}$/$^{e}n_{cr}$\tablenotemark{a} } 
&\colhead{$N_{mol}/I_{mol}$}  
&\colhead{$\frac{N(50)}{N(10)}$\tablenotemark{b}}  
\\
\colhead{Transition}  
&\colhead{\it (Dby)}
&\colhead{\it (GHz)}  
&\colhead{}
&\colhead{\it (K)}  
&\colhead{\it log(cm$^{-3}$)} 
&\colhead{\it ($\frac{cm^{2}}{K km s^{-1}}$ )} 
&\colhead{} 
}
\startdata 
SiO(2 -- 1; $v=1$)&3.098 &\nodata/21.712/\nodata & 2 &1787 &\nodata/\nodata &\nodata&\nodata  \\
SiO(2 -- 1; $v=0$)&3.098 &\nodata/21.712/\nodata & 2 &6.25 &5.27/5.66 &1.86(12)&2.94   \\
C$_{2}$H(1 -- 0; $\frac{3}{2}-\frac{1}{2}$)& 0.8&\nodata/43.675/\nodata &1 &4.19 &5.13/6.24 &2.32(13)&3.01 \\
C$_{2}$H(1 -- 0; $\frac{1}{2}-\frac{1}{2}$)& 0.8&\nodata/43.675/\nodata &1 &4.19 &5.13/6.24 &4.64(13)&3.01 \\
HCO${+}$(1 -- 0)& 3.888&\nodata/44.594/\nodata &1 & 4.28 & 5.15/5.80 & 9.51(11)&3.36  \\
HNC(1 -- 0)& 3.05&\nodata/45.332/\nodata &1 & 4.35 & 6.39/5.87& 1.51(12)&2.94   \\
HC$_{3}$N(10 -- 9)&3.72 &\nodata/4.5491/\nodata &10 &24.02 &5.86/6.87 &6.73(12)&0.723  \\
N$_{2}$H$^{+}$(1 -- 0)&3.40 &\nodata/46.587/\nodata &1 &4.47 &5.63/5.85&1.16(12) &3.00  \\
C$^{34}$S(2 -- 1)&1.96 &\nodata/24.104/\nodata &2 &6.94&5.65/6.17&4.06(12) &2.56  \\
CH$_{3}$OH(2$_{k}$ -- $1_{k}$)&0.89 &127.484/24.680/23.770 &4,3,4\tablenotemark{c} 
&6.98 &4.83/6.39&1.58(13) &5.96  \\
HNCO($5_{05}$ -- $4_{04}$)&1.60 &912.711/11.071/10.911 &5 &15.83 &6.29/5.99 &9.27(12) &3.15   \\
\enddata 
\tablecomments{Data from Splatalogue, the JPL Molecular Spectroscopy Catalog 
\citep[][]{JPLmol} and references within.  Collisional coefficients
are C$^{34}$S, HC$_{3}$N \citep[][]{GC78}, N$_{2}$H$^{+}$ \citep[][]{G75}, 
CH$_{3}$OH \citep[][]{PFD04}, HNCO \citep[][]{G86}, HCO$^{+}$ \citep[][]{F99} 
and SiO \citep[][]{DB06}.  For HNC and C$_{2}$H we adopt the collision 
coefficients of HCN \citep[][]{GT74}.}

\tablenotetext{a}{The critical density of the transitions neglecting
opacity effects ($^{H_{2}}n_{cr}\simeq \frac{A_{ul}}{C_{ul}}$) and
$^{e}n_{cr}$ based on the formalism of \citet[][]{DPGPR77}, with 
[e$^{-}] ~\simeq ~ 1.0 \times 10^{-5}$.}
\tablenotetext{b}{The ratio by which the derived column densities 
change when the $\rm T_{ex}$ is changed from 10 K $\rightarrow$ 
50 K.}
\tablenotetext{c}{$S_{ul}g_{K}g_{I}$ for the three blended $2_{1}-1_{1}$E, 
$2_{0}-1_{0}$E, and $2_{0}-1_{0}$A+ transitions \citep[eg.,][]{T91}.}
\end{deluxetable} 

\clearpage

\begin{figure}
\figurenum{2}
\epsscale{0.95}
\plotone{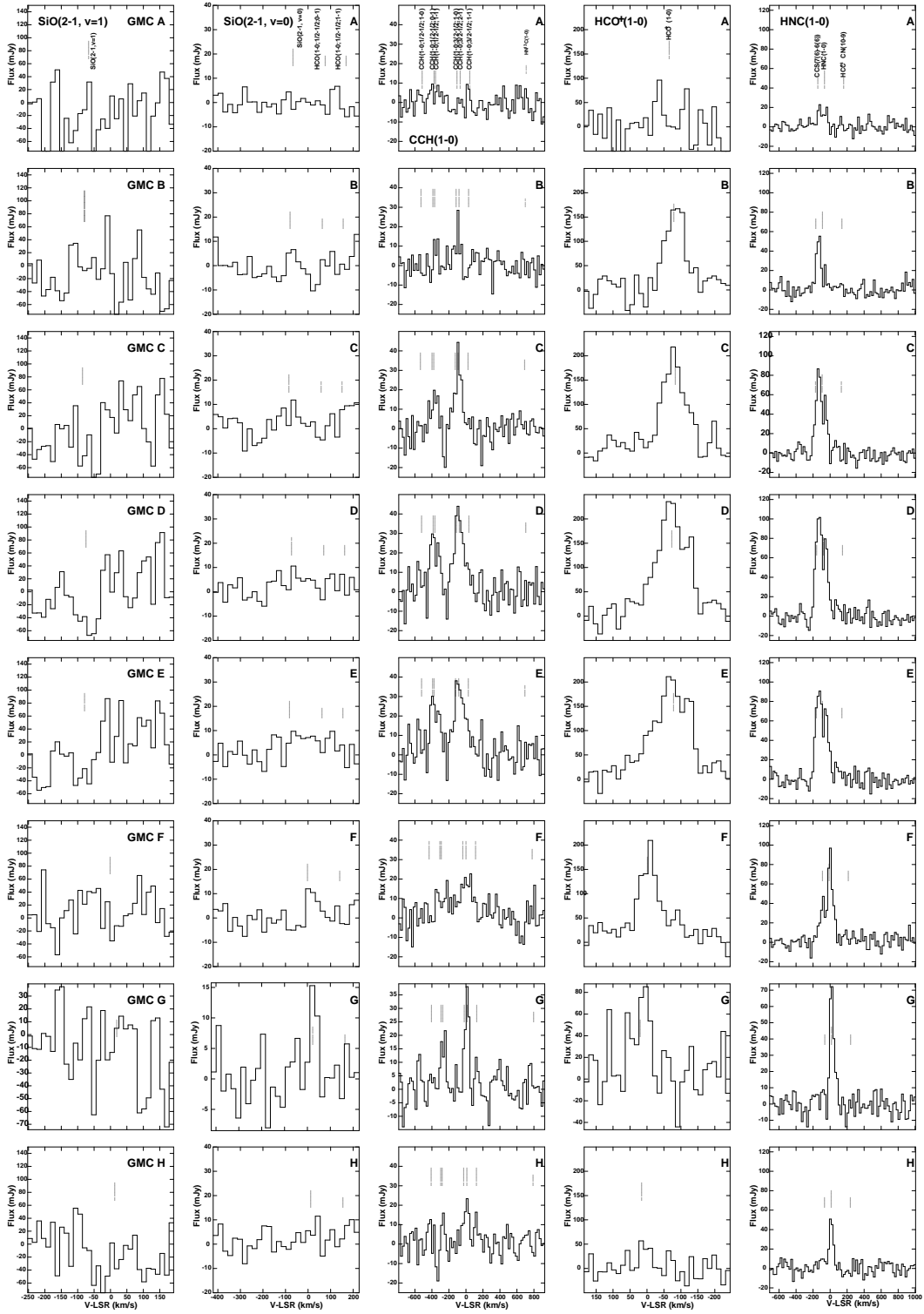}
\caption{Spectra of the molecular transitions towards Maffei
2's GMCs.  Spectra are summed over an 8$^{''}$ aperture centered on
the GMC positions indentified by \citet[][]{MTH08}. The vertical axis is the flux in 
the aperture and the x axis is LSR velocity.  \label{Fspec} }
\end{figure}
 
\clearpage

\begin{figure}
\figurenum{2}
\epsscale{0.95}
\plotone{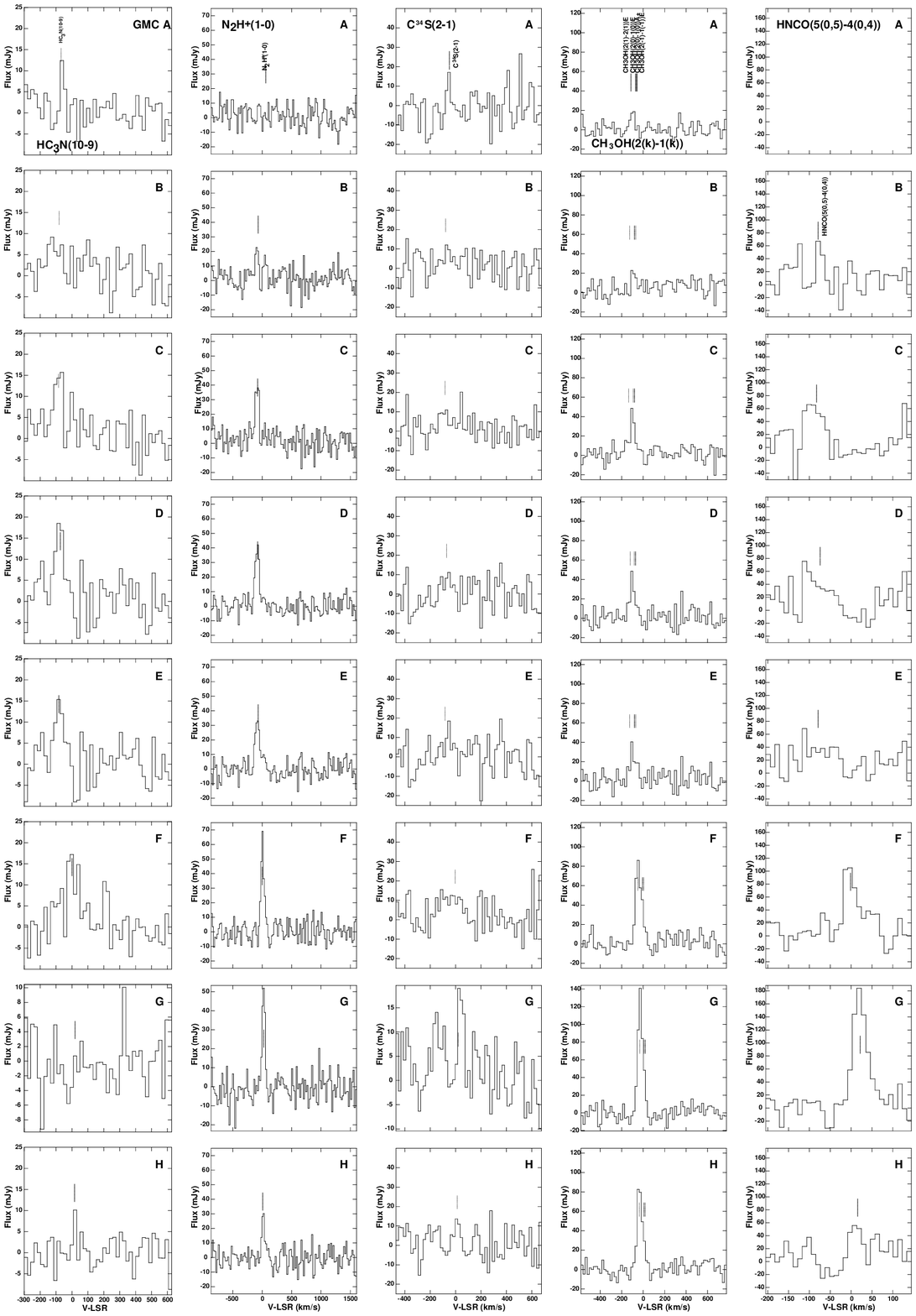}
\caption{Spectra of the molecular transitions towards Maffei
2's GMCs, continued.   \label{Fspec} }
\end{figure}
 
\clearpage

\begin{figure}
\figurenum{3}
\epsscale{0.95}
\plotone{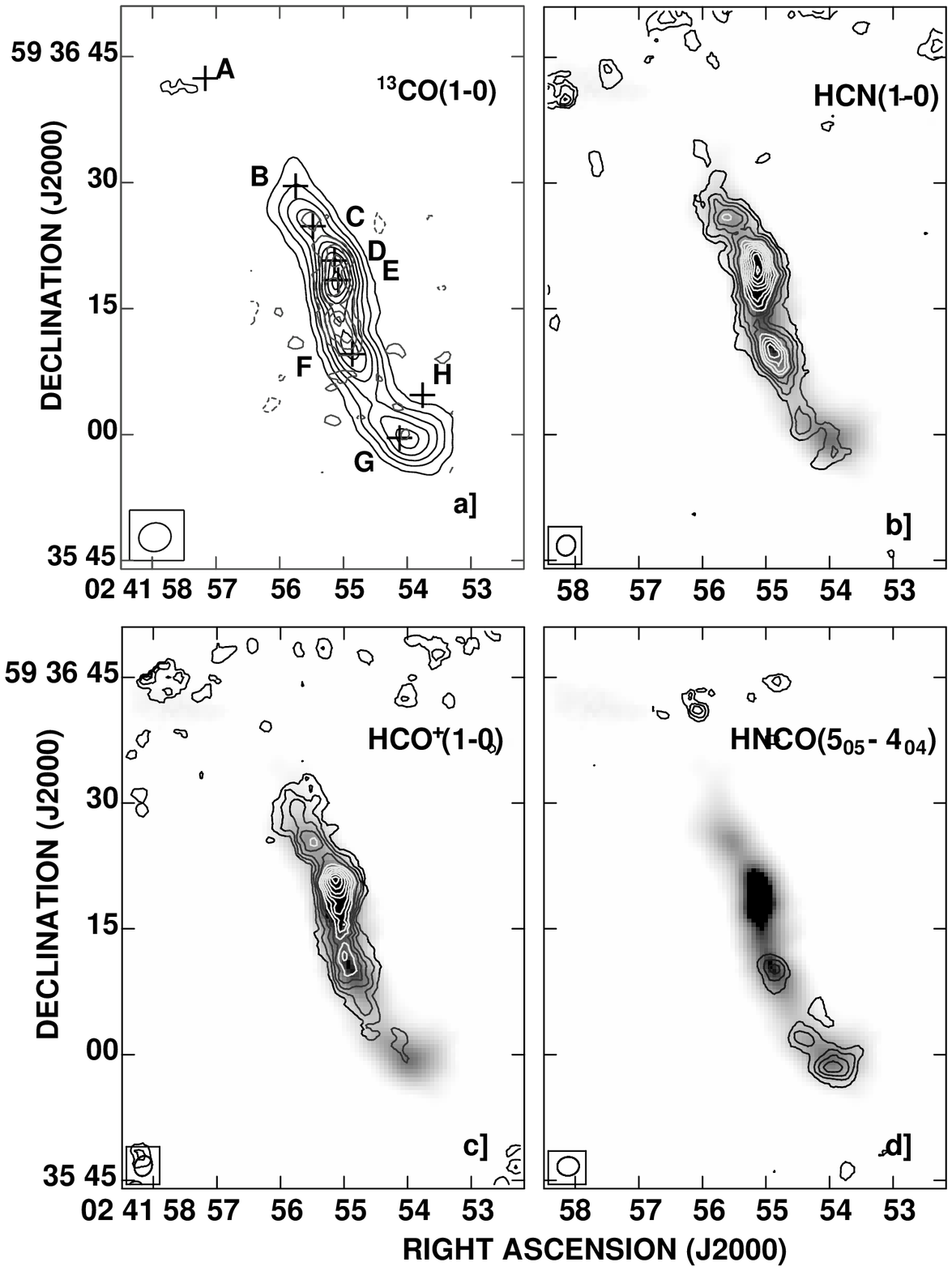}
\caption{Integrated intensity maps of four molecular transitions in
Maffei 2 observed with OVRO.  {\it a)} The $^{13}$CO(1--0) 
in steps of 0.75 Jy bm$^{-1}$ km s$^{-1}$. \citep[][]{MTH08} 
 {\it b)} The HCN(1--0) \citep[][]{MTH08} in steps of 0.50 Jy bm$^{-1}$
km s$^{-1}$ with a resolution of $2.5^{''}\times 2.2^{''}$.  {\it c)} 
The HCO$^{+}$(1--0) in steps of 0.50 Jy bm$^{-1}$
km s$^{-1}$ for a resolution given in Table \ref{ObsT}.  {\it d)} The
HNCO($5_{05}$--$4_{04}$) in steps of 0.40 Jy bm$^{-1}$ km s$^{-1}$.
In all planes the greyscale is the $^{13}$CO(1--0) integrated intensity. 
\label{FintiO} }
\end{figure}

\clearpage

\begin{figure}
\figurenum{4}
\epsscale{0.8}
\plotone{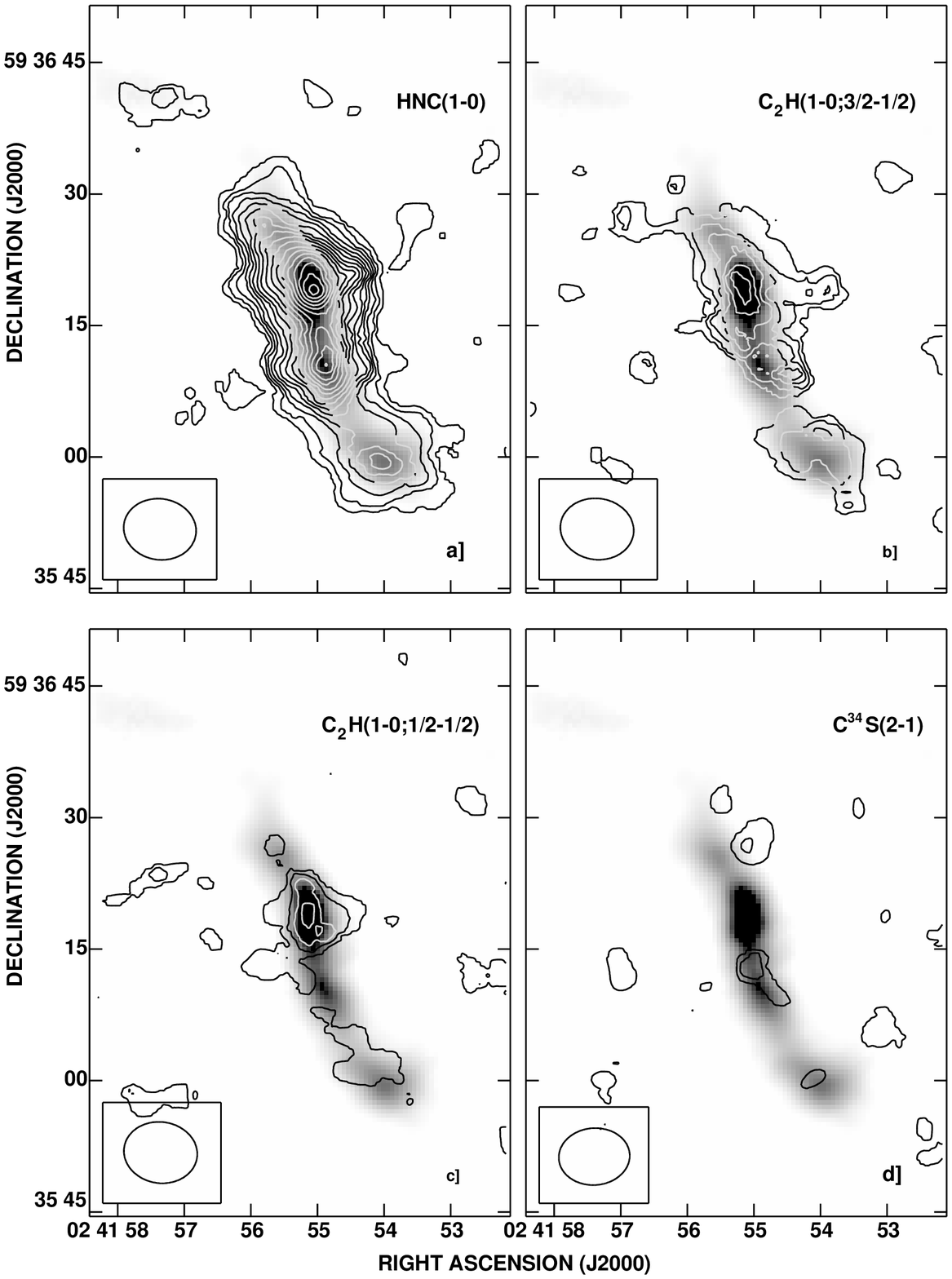}
\caption{Integrated intensity maps of the molecular transitions in
Maffei 2 observed with BIMA. {\it a)} The HNC(1--0) 
in steps of 1.0 Jy bm$^{-1}$ km s$^{-1}$.  {\it b)} The
C$_{2}$H(1--0;3/2--1/2) in steps of 1.1 Jy
bm$^{-1}$ km s$^{-1}$.  {\it c)} The C$_{2}$H(1--0;1/2--1/2)
in steps of 1.1 Jy bm$^{-1}$ km s$^{-1}$.  {\it
d)} The C$^{34}$S(2--1) in steps of 1.2 Jy
bm$^{-1}$ km s$^{-1}$.   In all planes the greyscale is the
$^{13}$CO(1--0) integrated intensity.  \label{FintiB} }
\end{figure}
 
\clearpage

\begin{figure}
\figurenum{4}
\epsscale{0.8}
\plotone{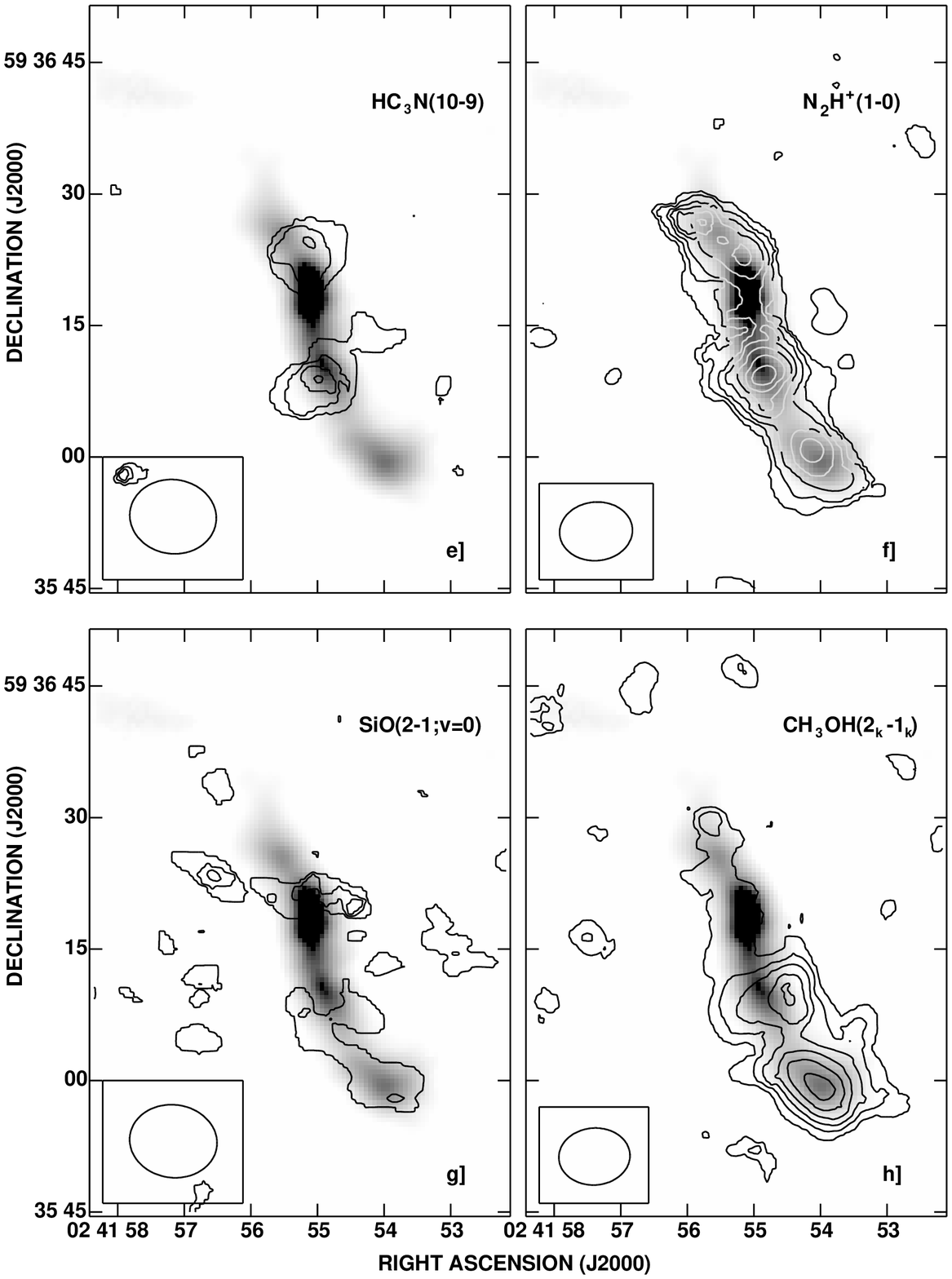}
\caption{Integrated intensity maps of the molecular transitions in
Maffei 2 observed with BIMA, continued. {\it e)} The HC$_{3}$N(10--9) 
integrated intensity in steps of 
levels 0.60 Jy bm$^{-1}$ km s$^{-1}$.  {\it f)} The
N$_{2}$H$^{+}$(1--0) in steps of 0.75 Jy
bm$^{-1}$ km s$^{-1}$.  {\it g)} The SiO(2--1; v=0) 
in steps of 0.40 Jy bm$^{-1}$ km s$^{-1}$.  {\it h)} The
CH$_{3}$OH(2$_{k}$--1$_{k}$) in steps of 1.5 Jy
bm$^{-1}$ km s$^{-1}$.  In all planes the greyscale is the
$^{13}$CO(1--0) integrated intensity.  \label{FintiB} }
\end{figure}
 
\clearpage

\begin{figure}
\figurenum{5}
\epsscale{0.95}
\plotone{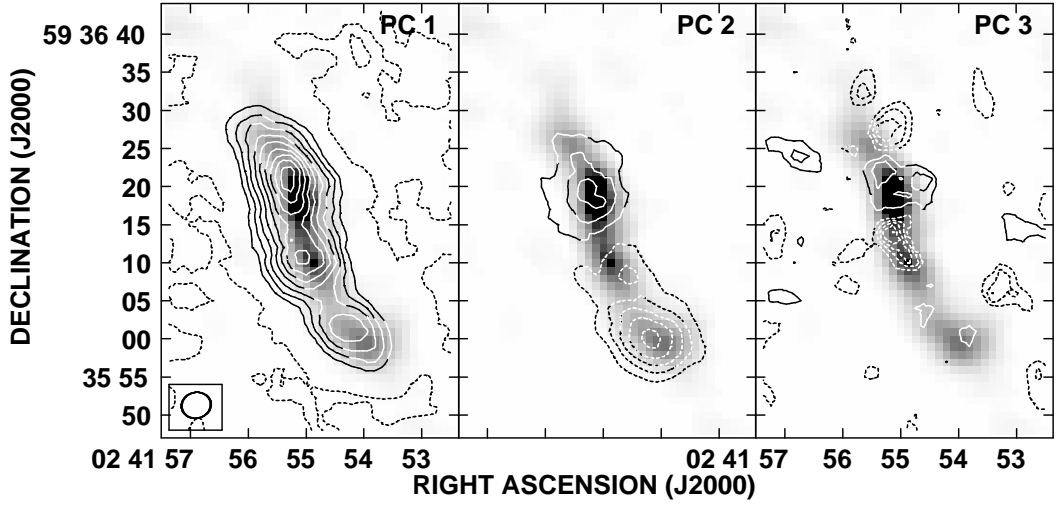}
\caption{Maps of the first three principal components of the molecular distribution.   
\label {Fpcamaps} }
\end{figure}

\begin{deluxetable}{lcccccccc} 
\tabletypesize{\scriptsize}
\tablenum{2} 
\tablewidth{0pt} 
\tablecaption{Observational Data\label{ObsT}} 
\tablehead{ 
\colhead{Transition}  
&\colhead{Dates}
&\colhead{Array}
&\colhead{Frequency}
&\colhead{T$_{sys}$}
&\colhead{$\Delta V_{chan}$} 
&\colhead{Beam}
&\colhead{K/Jy}  
&\colhead{Noise} \\ 
\colhead{}  
&\colhead{\it (MMYY)}
&\colhead{}
&\colhead{\it (GHz)}  
&\colhead{\it (K)}
&\colhead{($km~s^{-1}$)} 
&\colhead{\it ($^{''}\times^{''}$;$^{o}$)}  
&\colhead{}
&\colhead{\it (mJy bm$^{-1}$)}}  
\startdata 
BIMA: & & & & & & & & \\
SiO(2--1;v=0)\tablenotemark{a}& 0902-0303& B,C,D & 86.847 &140-1000 
& 21.57&$10.0\times 8.3$;83 & 1.68 & 10 \\ 
C$_{2}$H(1--0;3/2--1/2)\tablenotemark{a}& 0902-0303& B,C,D & 87.317 
&140-1000 & 21.46 &$8.4\times 7.0$;-84 & 2.75 & 11 \\ 
HNC(1--0)\tablenotemark{a}& 0902-0303& B,C,D & 90.664 &140-1000 
& 20.67 &$8.3\times 7.0$;81 & 2.59 & 11 \\ 
HC$_{3}$N(10--9)\tablenotemark{a}& 0902-0303& B,C,D & 90.979 & 140-1000 
& 20.60 &$9.9\times 8.5$;83 & 1.77 & 7.2 \\ 
N$_{2}$H$^{+}$(1--0)\tablenotemark{a}& 0902-0303& B,C,D & 93.174 &130-1100 
& 20.12 &$8.3\times 6.7$;-85 & 2.53 & 9.0 \\ 
C$^{34}$S(2--1)\tablenotemark{a}& 0902-0303& B,C,D & 96.413 &130-1100 
& 19.43 &$8.1\times 6.5$;-86 & 2.51 & 11 \\ 
CH$_{3}$OH($2_{k}$--$1_{k}$)\tablenotemark{a}& 0902-0303& B,C,D & 96.741 
&130-1100 & 19.37 &$8.1\times 6.5$;-86 & 2.48 & 15 \\
OVRO: & & & & & & & & \\
SiO(2--1;v=1)\tablenotemark{b}& 0199-0399& L,H,UH &86.243 &300-410 
& 13.90 &$2.3\times 2.0$;-56 & 36.5 & 9.5 \\ 
HCO$^{+}$(1--0)\tablenotemark{b}& 0199-0399& L,H,UH &89.189 &300-410 
& 13.44&$2.5\times 2.1$;-24  & 30.5& 6.7 \\ 
HNCO($5_{05}$--$4_{04}$)\tablenotemark{c}& 1098-0199& L,H & 109.906 &240-430 
& 10.92 &$2.6\times 2.2$;-84 & 17.3 & 6.3 \\ 
\enddata 
\tablenotetext{a}{Phase Center \#1: $\alpha = 02^{h} 41^{m} 55^{s}.0~~ 
\delta = +59^{o} 36' 15.^{''}0$ (J2000); v$_{lsr}$ = -15 km s$^{-1}$} 
\tablenotetext{b}{Phase Center \#1: $\alpha = 02^{h} 38^{m} 08^{s}.0~~ 
\delta = +59^{o} 23' 20.^{''}0$ (B1950); v$_{lsr}$ = -30 km s$^{-1}$ \\ 
$~~~~$Phase Center \#2: $\alpha = 02^{h} 38^{m} 08^{s}.25~~ 
\delta = +59^{o} 23' 27.^{''}0$ (B1950); v$_{lsr}$ = -30 km s$^{-1}$} 
\tablenotetext{c}{Phase Center \#1: $\alpha = 02^{h} 38^{m} 08^{s}.0~~ 
\delta = +59^{o} 23' 20.^{''}0$ (B1950); v$_{lsr}$ = -30 $km ~s^{-1}$ \\
$~~~~$Phase Center \#2: $\alpha = 02^{h} 38^{m} 08^{s}.25~~ 
\delta = +59^{o} 23' 27.^{''}0$ (B1950); v$_{lsr}$ = -30 $km ~s^{-1}$ } 
\end{deluxetable}

\clearpage 
 
\begin{deluxetable}{lccccc}
\footnotesize
\tablenum{3} 
\tablewidth{0pt} 
\tablecaption{Measured Intensities, Linewidths \& Centroids \label{IntT}} 
\tablehead{ 
\colhead{GMC} 
&\colhead{Transition}
&\colhead{I$_{mol}$} 
&\colhead{T$_{b}$} 
&\colhead{$v_{o}$}
&\colhead{$\Delta v$} \\
\colhead{$\alpha_{o}$,$\delta_{o}$} 
&\colhead{}
&\colhead{$(K~km~s^{-1})$}
&\colhead{$(mK)$}
&\colhead{$(km~s^{-1})$}  
&\colhead{$(km~s^{-1})$}
}
\startdata
A &SiO(1--0;v=1)& $<$19 & $<$88& \nodata & \nodata \\
(02:41:57.17)   &SiO(1--0;v=0)&$<$1.0 &$<$23 & \nodata & \nodata \\
(59:36:42.4)  &C$_{2}$H($\frac{3}{2}$--$\frac{1}{2}$)&$<$3.3 &  $<$34& \nodata & \nodata \\
  &C$_{2}$H($\frac{1}{2}$--$\frac{1}{2}$)&$<$3.3 &  $<$34& \nodata & \nodata \\
  &HCO$^{+}$(1--0) & 16$\pm$7  & 270$\pm$38 & -34$\pm$5 & 21$\pm$10 \\
  &HNC(1--0) &  $<$2.9  & 45$\pm$16 & -100$\pm$11 & 120$\pm$25 \\
  &HC$_{3}$N(10--9) & $<$1.0  & $<$26 & \nodata &\nodata \\
  &N$_{2}$H$^{+}$(1--0) & $<$1.9  & $<$30 & \nodata &\nodata \\
  &C$^{34}$S(2--1) & $<$2.2  & $<$28 & \nodata &\nodata \\
  &CH$_{3}$OH($2_{k}$--$1_{k}$) & $<$3.2  & 46$\pm$19 & -100$\pm$6 &45$\pm$15 \\
  &HNCO($5_{05}$--$4_{04}$) & \nodata  & \nodata & \nodata &\nodata \\
B &SiO(1--0;v=1)& $<$19 & $<$88& \nodata & \nodata \\
(02:41:55.75)   &SiO(1--0;v=0)&$<$1.0 &35$\pm$12& -67$\pm$13 & 55$\pm$30 \\
(59:36:29.6)  &C$_{2}$H($\frac{3}{2}$--$\frac{1}{2}$)&$<$3.3 & $<$34& \nodata & \nodata \\
  &C$_{2}$H($\frac{1}{2}$--$\frac{1}{2}$)&$<$3.3 &  $<$34& \nodata & \nodata \\
  &HCO$^{+}$(1--0) & 61$\pm$7  & 470$\pm$38 & -84$\pm$3 & 68$\pm$7 \\
  &HNC(1--0) & 9.1$\pm$1 & 150$\pm$16  & -134$\pm$3 & 65$\pm$7  \\
  &HC$_{3}$N(10--9) & $<$1.0  & $<$26 & \nodata &\nodata \\
  &N$_{2}$H$^{+}$(1--0) & 5.1$\pm$1  & 60$\pm$15 & -98$\pm$6 &50$\pm$10 \\
  &CH$_{3}$OH($2_{k}$--$1_{k}$) & 10$\pm$2  & 53$\pm$19 & -87$\pm$8 & 48$\pm$20 \\
  &C$^{34}$S(2--1) & $<$2.2  & $<$28 & \nodata &\nodata \\
  &HNCO($5_{05}$--$4_{04}$) & $\lsim$4.8 & 50$\pm$29 & -112$\pm$20 & 110$\pm$50 \\
 C &SiO(1--0;v=1)& $<$19 & $<$88& \nodata & \nodata \\
(02:41:55.48)   &SiO(1--0;v=0)&$<$1.0 &24$\pm$12& -80$\pm$20 & 83$\pm$47 \\
(59:36:24.8)  &C$_{2}$H($\frac{3}{2}$--$\frac{1}{2}$)&11$\pm$2 & 100$\pm$17& -88$\pm$6 & 93$\pm$15 \\
  &C$_{2}$H($\frac{1}{2}$--$\frac{1}{2}$)&3.9$\pm$2 & 48$\pm$17& -76$\pm$17 & 90$\pm$40 \\
  &HCO$^{+}$(1--0) & 82$\pm$7  & 520$\pm$38 & -78$\pm$3 & 83$\pm$6 \\
  &HNC(1--0) & 24$\pm$1 & 200$\pm$16  & -120$\pm$4 & 120$\pm$8  \\
  &HC$_{3}$N(10--9) & 2.8$\pm$0.5  & 42$\pm$13 & -88$\pm$7 & 65$\pm$16 \\
  &N$_{2}$H$^{+}$(1--0) & 9.6$\pm$1  & 110$\pm$15 & -88$\pm$5 &77$\pm$10 \\
  &CH$_{3}$OH($2_{k}$--$1_{k}$) & 5.0$\pm$2  & 100$\pm$19 & -99$\pm$4 & 48$\pm$8 \\
  &C$^{34}$S(2--1) & $<$2.2  & $<$28 & \nodata &\nodata \\
  &HNCO($5_{05}$--$4_{04}$) & $\lsim$4.8 & 130$\pm$29 & -89$\pm$5 & 42$\pm$10 \\
 D &SiO(1--0;v=1)& $<$19 & $<$88& \nodata & \nodata \\
(02:41:55.14)   &SiO(1--0;v=0)&2.5$\pm$0.5 &$<$23& \nodata & \nodata \\
(59:36:20.7)  &C$_{2}$H($\frac{3}{2}$--$\frac{1}{2}$)&17$\pm$2 & 110$\pm$17& -82$\pm$9 & 130$\pm$22 \\
  &C$_{2}$H($\frac{1}{2}$--$\frac{1}{2}$)&11$\pm$2 & 80$\pm$17& -77$\pm$17 & 120$\pm$35 \\
  &HCO$^{+}$(1--0) & 180$\pm$7  & 580$\pm$38 & -74$\pm$3 & 110$\pm$7 \\
  &HNC(1--0) & 41$\pm$1 & 250$\pm$16  & -109$\pm$4 & 140$\pm$8  \\
  &HC$_{3}$N(10--9) & 3.2$\pm$0.5  & 50$\pm$13 & -80$\pm$6 & 67$\pm$15 \\
  &N$_{2}$H$^{+}$(1--0) & 9.4$\pm$1  & 110$\pm$15 & -87$\pm$4 &82$\pm$8 \\
  &CH$_{3}$OH($2_{k}$--$1_{k}$) & 5.5$\pm$2  & 95$\pm$19 & -101$\pm$5 & 57$\pm$12 \\
  &C$^{34}$S(2--1) & $<$2.2  & $<$28 & \nodata &\nodata \\
  &HNCO($5_{05}$--$4_{04}$) & $\lsim$4.8 & 86$\pm$29 & -90$\pm$11 & 82$\pm$15 \\
  E &SiO(1--0;v=1)& $<$19 & $<$88& \nodata & \nodata \\
(02:41:55.08)   &SiO(1--0;v=0)& $\sim$0.77 &22$\pm$12& -2.0$\pm$26  & 220$\pm$60 \\
(59:36:18.4)  &C$_{2}$H($\frac{3}{2}$--$\frac{1}{2}$)&19$\pm$2 & 97$\pm$17& -73$\pm$12 & 160$\pm$28 \\
  &C$_{2}$H($\frac{1}{2}$--$\frac{1}{2}$)&13$\pm$2 & 77$\pm$17& -72$\pm$16 & 130$\pm$38 \\
  &HCO$^{+}$(1--0) & 130$\pm$7  & 520$\pm$38 & -73$\pm$4 & 130$\pm$8 \\
  &HNC(1--0) & 44$\pm$1 & 220$\pm$16  & -104$\pm$4 & 150$\pm$10  \\
  &HC$_{3}$N(10--9) & 2.1$\pm$0.5  & 39$\pm$13 & -80$\pm$8 & 67$\pm$18 \\
  &N$_{2}$H$^{+}$(1--0) & 6.6$\pm$1  & 85$\pm$15 & -83$\pm$5 &92$\pm$12 \\
  &CH$_{3}$OH($2_{k}$--$1_{k}$) & 6.7$\pm$2  & 70$\pm$19 & -94$\pm$8 & 75$\pm$18 \\
  &C$^{34}$S(2--1) & $<$2.2  & $<$28 & \nodata &\nodata \\
  &HNCO($5_{05}$--$4_{04}$) & $<$4.8 & $<$58 & \nodata & \nodata \\
  F &SiO(1--0;v=1)& $<$19 & $<$88& \nodata & \nodata \\
(02:41:54.86)   &SiO(1--0;v=0)& $<$1.0 &35$\pm$12& 17$\pm$8  & 50$\pm$18 \\
(59:36:09.6)  &C$_{2}$H($\frac{3}{2}$--$\frac{1}{2}$)&9.4$\pm$2 & 63$\pm$17& 10$\pm$12 & 130$\pm$50 \\
  &C$_{2}$H($\frac{1}{2}$--$\frac{1}{2}$)&$<$3.3 & $\sim$46& \nodata & \nodata \\
  &HCO$^{+}$(1--0) & 85$\pm$7  & 450$\pm$38 & -6.9$\pm$4 & 83$\pm$8 \\
  &HNC(1--0) & 36$\pm$1 & 250$\pm$16  & -0.9$\pm$3 & 67$\pm$17  \\
  &HC$_{3}$N(10--9) & 4.1$\pm$0.5  & 31$\pm$13 & 0.1$\pm$13 & 130$\pm$30 \\
  &N$_{2}$H$^{+}$(1--0) & 15$\pm$1  & 150$\pm$15 & 5.4$\pm$6 &70$\pm$10 \\
  &CH$_{3}$OH($2_{k}$--$1_{k}$) & 15$\pm$2  & 190$\pm$19 & -43$\pm$3 & 68$\pm$5 \\
  &C$^{34}$S(2--1) & 5.0$\pm$1  & $\sim$30 & -46$\pm$26 &180$\pm$62 \\
  &HNCO($5_{05}$--$4_{04}$) & 22$\pm$2 & 190$\pm$29 & -1.6$\pm$3 & 40$\pm$7 \\
\enddata
\end{deluxetable}

\clearpage

\begin{deluxetable}{lccccc}
\footnotesize
\tablenum{3} 
\tablewidth{0pt} 
\tablecaption{Measured Intensities, Linewidths \& Centroids \label{IntT}} 
\tablehead{ 
\colhead{GMC} 
&\colhead{Transition}
&\colhead{I$_{mol}$} 
&\colhead{T$_{b}$} 
&\colhead{$v_{o}$}
&\colhead{$\Delta v$} \\
\colhead{$\alpha_{o}$,$\delta_{o}$} 
&\colhead{}
&\colhead{$(K~km~s^{-1})$}
&\colhead{$(mK)$}
&\colhead{$(km~s^{-1})$}  
&\colhead{$(km~s^{-1})$}
}
\startdata
 G &SiO(1--0;v=1)& $<$19 & $<$88& \nodata & \nodata \\
(02:41:54.12)   &SiO(1--0;v=0)& 1.2$\pm$0.5 &52$\pm$17& 28$\pm$4  & 32$\pm$13 \\
(59:35:59.6)  &C$_{2}$H($\frac{3}{2}$--$\frac{1}{2}$)&6.1$\pm$2 & 110$\pm$17& 12$\pm$4 & 55$\pm$10 \\
  &C$_{2}$H($\frac{1}{2}$--$\frac{1}{2}$)&4.1$\pm$2 & 51$\pm$17& 27$\pm$12 & 72$\pm$ 28\\
  &HCO$^{+}$(1--0) & $\sim$15$\pm$7  & 200$\pm$38 & 15$\pm$7 & 62$\pm$15 \\
  &HNC(1--0) & 17$\pm$1 & 200$\pm$16  & 20$\pm$2 & 55$\pm$5  \\
  &HC$_{3}$N(10--9) & $<$1.5  & $<$26 & \nodata & \nodata \\
  &N$_{2}$H$^{+}$(1--0) & 10$\pm$1  & 140$\pm$15 & 23$\pm$3 &57$\pm$7 \\
  &CH$_{3}$OH($2_{k}$--$1_{k}$) & 25$\pm$2  & 300$\pm$19 & -27$\pm$2 & 55$\pm$3 \\
  &C$^{34}$S(2--1) & 3.3$\pm$2  & 47$\pm$19 & 40$\pm$6 &50$\pm$15 \\
  &HNCO($5_{05}$--$4_{04}$) & 33$\pm$2 & 320$\pm$29 & 21$\pm$2 & 43$\pm$3 \\
 H &SiO(1--0;v=1)& $<$19 & $<$88& \nodata & \nodata \\
(02:41:53.76)  &SiO(1--0;v=0)& $<$1.0 & $<$23& \nodata & \nodata \\
(59:36:04.7)  &C$_{2}$H($\frac{3}{2}$--$\frac{1}{2}$)&$<$3.3 & 60$\pm$17& 5.8$\pm$9 & 78$\pm$22 \\
  &C$_{2}$H($\frac{1}{2}$--$\frac{1}{2}$)&$<$3.3 & $<$34& \nodata & \nodata \\
  &HCO$^{+}$(1--0) & $<$15  & 150$\pm$38 & 6.2$\pm$6 & 35$\pm$13 \\
  &HNC(1--0) & 7.8$\pm$1 & 130$\pm$16  & 10$\pm$5 & 58$\pm$17  \\
  &HC$_{3}$N(10--9) & $<$1.0  & $<$26 & \nodata & \nodata \\
  &N$_{2}$H$^{+}$(1--0) & $<$1.9  & 73$\pm$15 & 20$\pm$6 &57$\pm$10 \\
  &CH$_{3}$OH($2_{k}$--$1_{k}$) & 8.4$\pm$2  & 200$\pm$19 & -32$\pm$2 & 60$\pm$5 \\
  &C$^{34}$S(2--1) & $<$2.2  & $<$28 & \nodata &\nodata \\
  &HNCO($5_{05}$--$4_{04}$) & $\lsim$4.8 & 110$\pm$29 & 11$\pm$4 & 32$\pm$10 \\     
\enddata
\tablecomments{T$_{b}$ is the main-beam brightness temperature 
based on a 8$\arcsec$ aperture.   $I_{mol}$ is the peak integrated intensity 
in units of K km s$^{-1}$ for the resolution given in Table \ref{ObsT}.  
Uncertainties are based on the RMS noise for the temperatures and intensities, 
and $1\sigma$ from the least-squared gaussian fits for the velocity information.  
Upper limits represent 2$\sigma$ values.}
\end{deluxetable}

\begin{figure}
\figurenum{6}
\epsscale{0.95}
\plotone{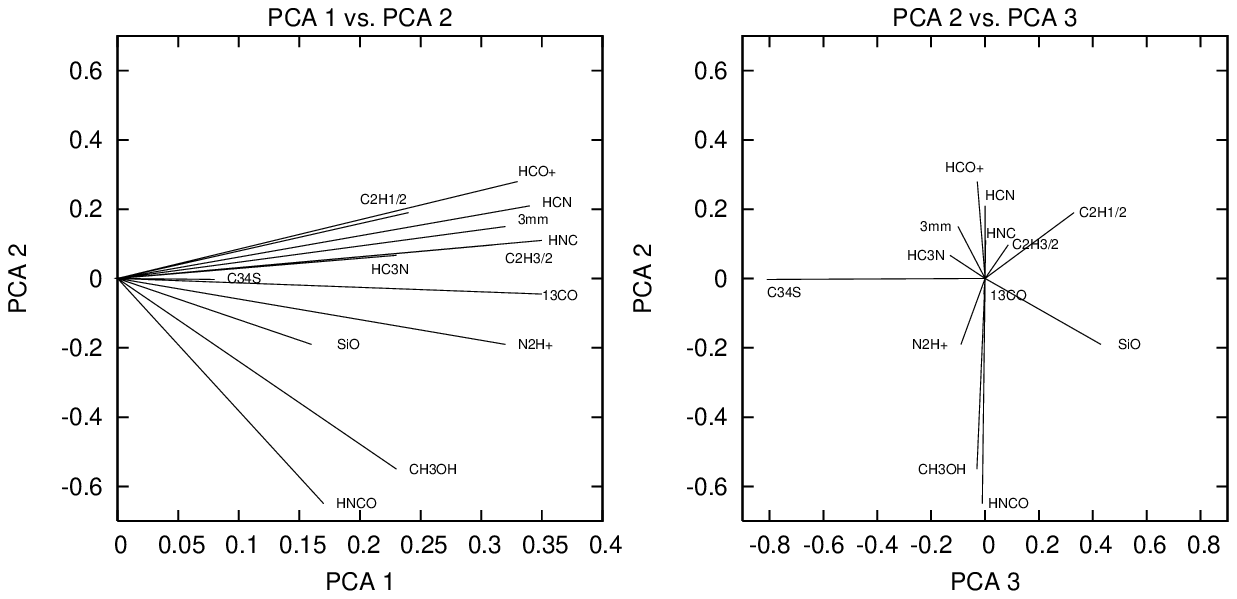}
\caption{Projections of each map onto the first three principal component maps. {\it Left)} 
Projections onto the first and second principal components.  {\it Right)} Projections onto the 
second and third component, while the figure plane geometry is such that it may be seen 
as a view looking to the left down the PC 1 axis. \label{Fpcavect} }
\end{figure}
 
\clearpage
 
\begin{figure}
\figurenum{7}
\epsscale{0.8}
\plotone{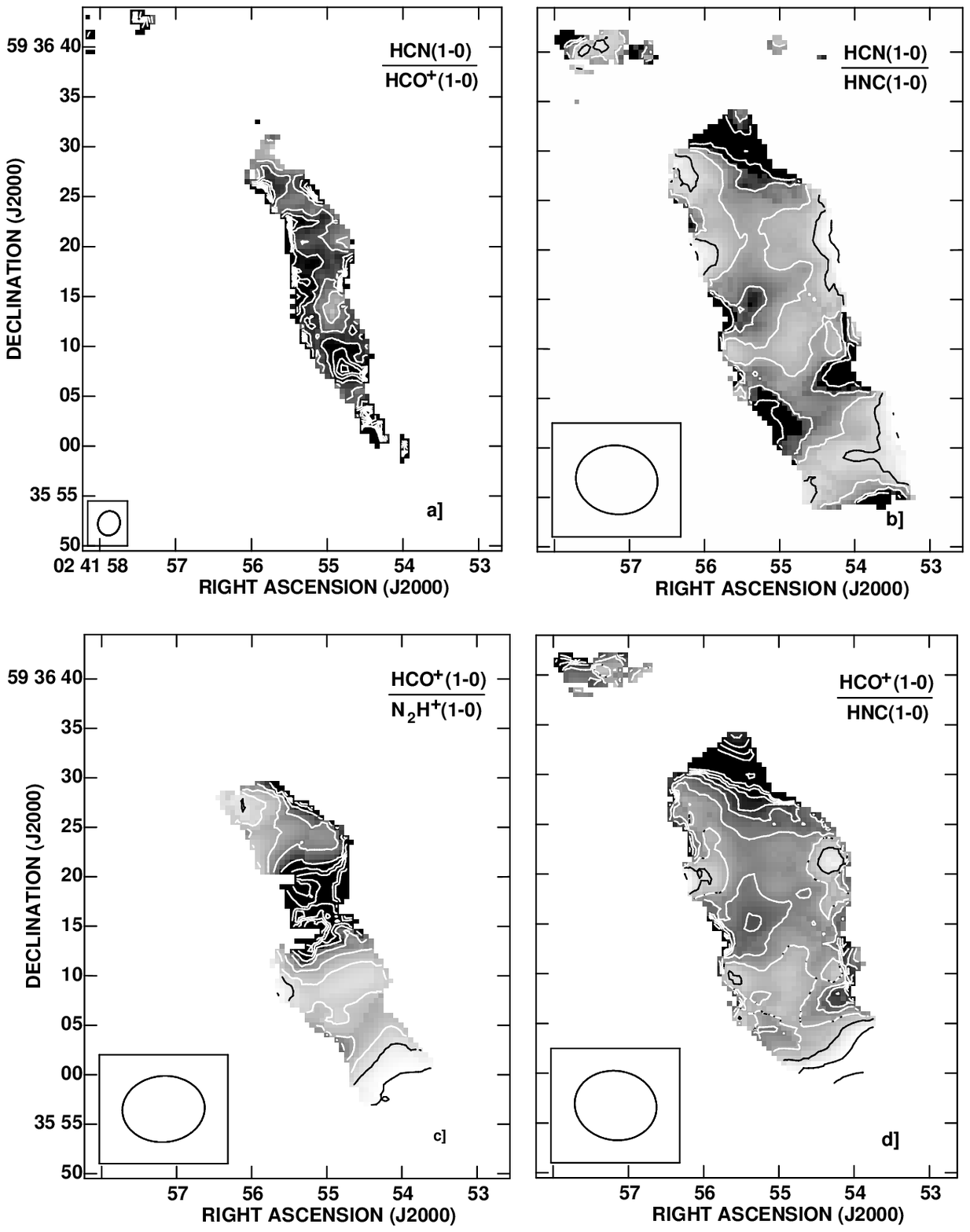}
\caption{{\it a)} The HCN(1--0)/HCO$^{+}$(1--0) line ratio.  The
  greyscale ranges from 0 to 1.5 with dark being large ratios and
  contours are in steps of 0.333 for the resolution of the HNC(1--0)
  dataset.  HCN(1--0) is from \citet[][]{MTH08}. {\it b)} The
  HCN(1--0)/HNC(1--0) line ratio with the greyscale ranging from  
  0.5 to 2.0 and contours are in steps of
  0.45 for the resolution of the HNC(1--0) dataset.  {\it c)} The
  HCO$^{+}$(1--0)/N$_{2}$H$^{+}$(1--0) line ratio  with the greyscale ranging from
  1.0 to 6.0 and contours are in steps of 1.0. {\it d)} The
  HCO$^{+}$(1--0)/HNC(1--0) line ratio  with the greyscale ranging from
  0.5 to 2.0 and contours are in steps of  0.25. 
\label{Fdenserat} }
\end{figure}

\begin{figure}
\figurenum{8}
\epsscale{0.95}
\plotone{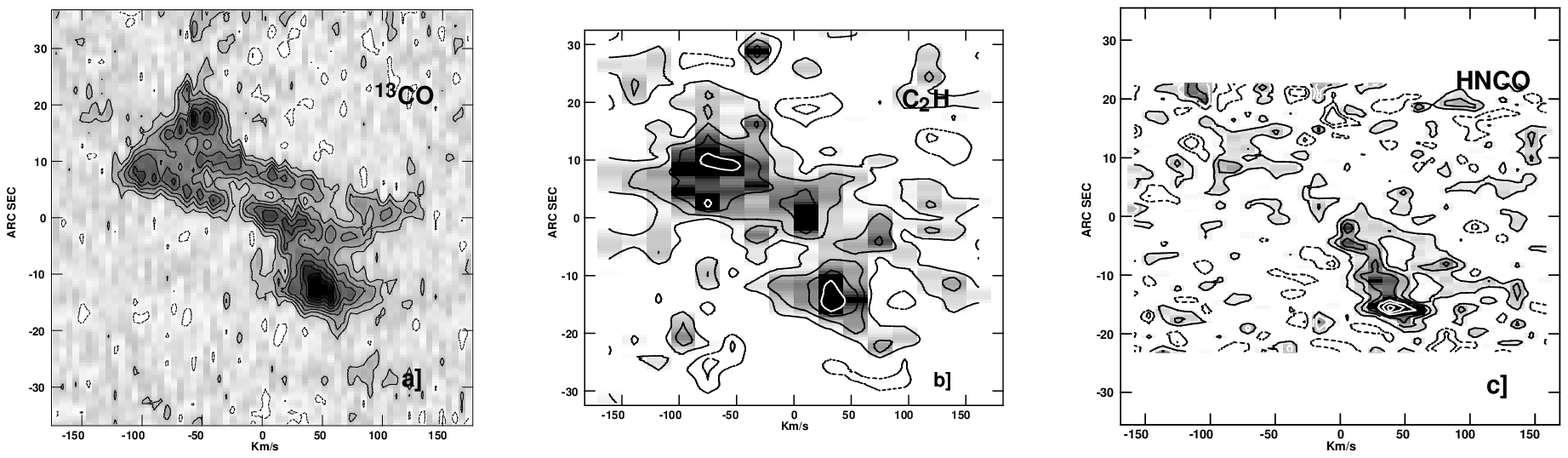}
\caption{The Position-Velocity diagrams for {\it a)} $^{13}$CO(1--0) 
\citep[][]{MTH08}, {\it b)} CCH(1--0; 3/2--1/2), and {\it c)} 
HNCO($5_{05}$--$4_{04}$). The reference locations for each map are within 
$\pm$1$^{''}$ of 02$^{h}$41$^{m}$55$^{s}$, 59$^{o}$36$^{'}$10$^{''}$ and $\pm$2 
km s$^{-1}$ of -20 km s$^{-1}$. The position angle is taken at 29$^{o}$. \label{Fpv} }
\end{figure}
 
\clearpage

\begin{figure}
\figurenum{9}
\epsscale{0.45}
\plotone{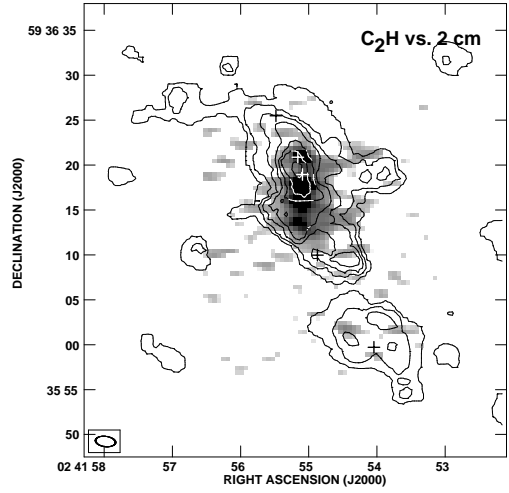}
\caption{C$_{2}$H contours overlaid on the 2 cm radio continuum \citep[][]{TH94}. Contours 
are the same as in Fig. \ref{FintiO} and the greyscale ranges from  0.3 mJy bm$^{-1}$ 
to 2 mJy bm$^{-1}$ for a beam of $2.^{''}2\times 1.^{''}2$.\label{cchoutf} }
\end{figure}

\begin{figure}
\figurenum{10}
\epsscale{0.35}
\plotone{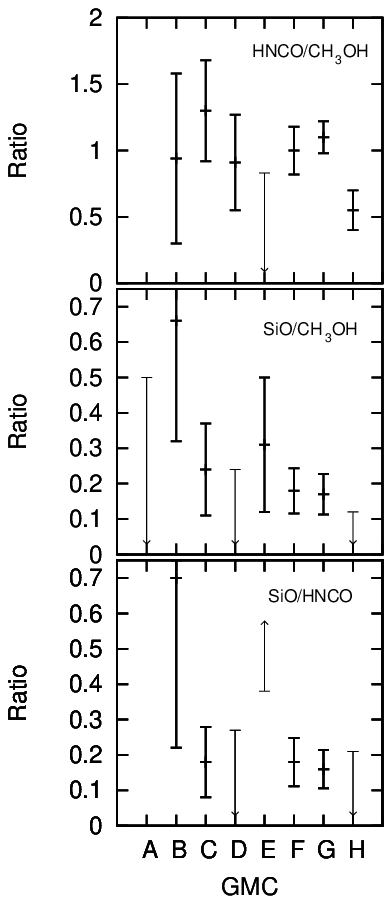}
\caption{The shock tracer peak antenna temperature ratios. Ratios are measured over an 
8$^{''}$ aperture centered on each cloud.  Downward (upward) facing arrows are upper (lower) 
limits. \label{hncoch3oh} }
\end{figure}

\clearpage

\begin{figure}
\figurenum{11}
\epsscale{0.75}
\plotone{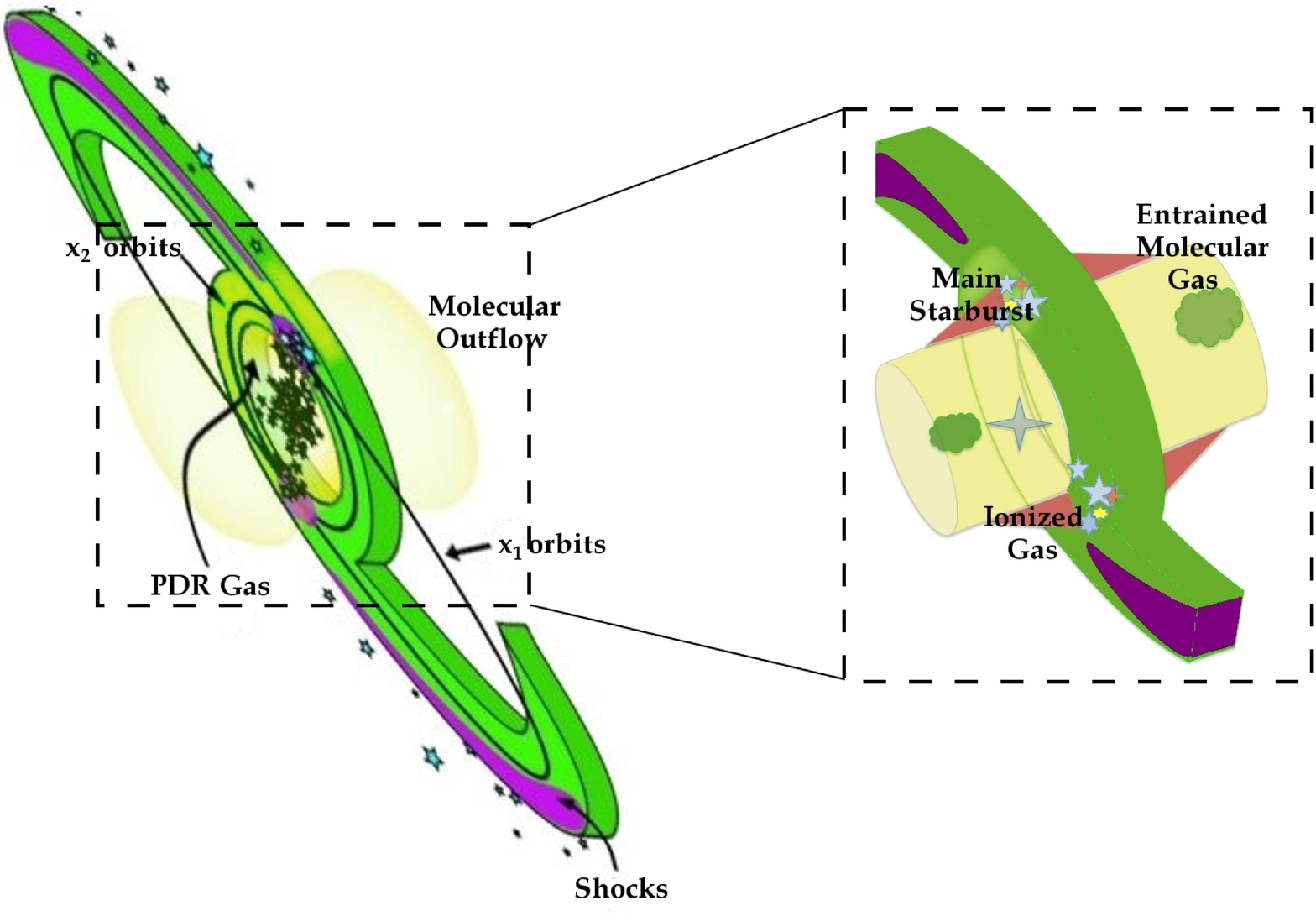}
\caption{Schematic of the overall chemistry of Maffei 2.  {\it Left)} The schematic adapted from 
Fig. 10 of \citet[][]{MT05} to emphasize the similarity between the two nuclei's chemical properties.  
{\it Right)} A zoom in of the central molecular ring region with its star formation and molecular 
outflow.  \label{Fgenchem} }
\end{figure}

\begin{deluxetable}{lccccccccccc}
\tabletypesize{\scriptsize}
\tablenum{4} 
\tablewidth{0pt} 
\tablecaption{Molecular Abundances in Maffei 2 \label{AbuT}} 
\tablehead{ 
\colhead{}  
&\colhead{N(H$_{2}$)}
&\colhead{SiO} 
&\colhead{C$_{2}$H}
&\colhead{C$_{2}$H$_{cor}$\tablenotemark{a}}
&\colhead{HCO$^{+}$} 
&\colhead{HNC}
&\colhead{HC$_{3}$N}   
&\colhead{N$_{2}$H$^{+}$} 
&\colhead{C$^{34}$S}  
&\colhead{CH$_{3}$OH}
&\colhead{HNCO}  
} 
\startdata
A  &2.8(21) & $<$6.8(-10)&$<$5.4(-8) &\nodata& 5.4(-9) & $<$1.6(-9) & 
$<$2.4(-9) & $<$7.9(-10) & $<$3.2(-9) & $<$1.8(-8) & \nodata \\
B  &9.7(21) & $<$2.0(-10) & $<$1.5(-8) &\nodata & 6.1(-9) & 1.4(-9) & 
$<$6.9(-10) & 6.1(-10) & $<$9.2(-10) & 1.6(-8) & $<$1.6(-8) \\
C  &1.9(22) & $<$1.0(-10) & 1.8(-8) &1.8(-8) & 4.2(-9) & 1.9(-9) & 
1.0(-9) & 5.8(-10) & $<$4.9(-10) & 4.2(-9) & $<$2.4(-8) \\
D  &2.9(22) & 1.6(-10) & 2.2(-8) &4.2(-8)& 6.0(-9) & 2.1(-9) & 
7.6(-10) & 3.8(-10) & $<$3.0(-11)  & 3.0(-9) & $\sim$1.1(-9) \\
E  &3.0(22) & $\sim$4.7(-11) & 2.5(-8) &4.8(-8) & 4.1(-9) & 2.2(-9) & 
4.7(-10) &2.6(-10) & $<$3.0(-10) &  3.7(-9) & $<$1.5(-9) \\
F &2.3(22) & $\lsim$8.3(-11) & 9.6(-9)&9.6(-9)& 3.5(-9) & 2.3(-9) & 
1.2(-9)& 7.4(-10) & 8.7(-10) & 1.0(-8) & 8.7(-9) \\
G &2.0(22) & $\sim$1.1(-10) & 1.2(-8)& 2.3(-8) & 7.2(-10) & 1.3(-9) & 
$<$5.0(-10) & 6.0(-10) & $\sim$6.5(-10) & 2.0(-8) & 1.6(-8) \\
H &8.9(21) & $<$2.1(-10) & $<$1.7(8)& $<$2.0(-9) & $<$1.6(-9) & 1.3(-9) & 
$<$7.5(-10) & $<$2.5(-10) & $<$1.0(-9) & 1.5(-8) & $<$5.1(-9) \\
\enddata
\tablecomments{Format for entries are a(b)= a$\times 10^{b}$.  Each
molecule is based on the assumptions of optically thin line emission
with T$_{ex}$ also 10 K.  Upper limits are 2$\sigma$.  Uncertainties
are dominated by systematics and are at least a factor of 3 (see text
for discussion of uncertainties).  H$_{2}$ column densities are based
on $^{13}$CO(1-0) emission sampled at the resolution of the HNC 
data with an abundance of [H$_{2}$/$^{13}$CO]  = 
7.06$\times 10^{5}$ adopted \citep[][]{MTH08}.}
\tablenotetext{a}{Corrected for finite optical depth based of the 
fine structure line ratios (Table \ref{RatT}).}
\end{deluxetable}

\begin{deluxetable}{lcccc}
\tablenum{5} 
\tablewidth{0pt} 
\tablecaption{Other Observed Transitions \label{othermolT}} 
\tablehead{ 
\colhead{Molecule}
&\colhead{Transition}
& \colhead{$\nu$} 
& \colhead{T$_{mb}$} 
& \colhead{GMCs} \\  
\colhead{}
&\colhead{}
& \colhead{$(GHz)$} 
& \colhead{$(mJy/bm)$} 
& \colhead{}
}
\startdata
HCO &  $1_{0,1}$ -- $0_{0,0}$; $\frac{1/2}{1/2}$;  1 -- 1 & 86.777& $<$20 & \nodata \\
HCO &  $1_{0,1}$ -- $0_{0,0}$; $\frac{1/2}{1/2}$;  0 -- 1 & 86.806& $<$20 & \nodata \\
HN$^{13}$C &  1 -- 0 & 87.091& $<$22 & \nodata \\
HCC$^{13}$CN/HC$^{13}$C$_{2}$N&  10 -- 9 & 90.593/90.602& $<$22 & \nodata \\
\nodata \\
\hline
\nodata &\nodata& 89.222(5) & 70$\pm$15 & C \\
\nodata&\nodata& 90.912(5) & 10$\pm$6 & F \\
\enddata
\tablecomments{Upper limits are 2$\sigma$.}
\end{deluxetable}

\begin{deluxetable}{lcccccc}
\footnotesize
\tablenum{6} 
\tablewidth{0pt} 
\tablecaption{Selected Intensities Ratios \label{RatT}} 
\tablehead{ 
\colhead{Location} 
& \colhead{$\frac{HCN(1-0)}{HNC(1-0)}$}
& \colhead{$\frac{CCH(3/2-1/2)}{CCH(1/2-1/2)}$}
& \colhead{$\frac{HCN(1-0)}{HCO^{+}(1-0)}$}
& \colhead{$\frac{HNC(1-0)}{HN^{13}C(1-0)}$}
& \colhead{$\frac{CS(2-1)}{C^{34}S(2-1)}$}
& \colhead{$\frac{HCN(1-0)}{CS(2-1)}$}
}
\startdata
A & \nodata & \nodata & \nodata &$>$0.79& $>$1.4 &$\lsim$1.6 \\
B & 2.4$\pm$0.4 & \nodata & 0.63$\pm$0.1 & $>$4.5 & $>$1.4 &4.9$\pm$1.5 \\
C & 1.6$\pm$0.2 & 2.8$\pm$1.5 & 0.90$\pm$0.1 & $>$6.6& $>$6.0 & 2.5$\pm$0.8\\
D & 1.3$\pm$0.2 & 1.5$\pm$0.3 & 1.1$\pm$0.2 &$>$10& $>$14 & 2.1$\pm$0.6\\
E & 1.3$\pm$0.2 & 1.5$\pm$0.3 & 1.3$\pm$0.2 &$>$9.4&$>$13 & 2.2$\pm$0.7\\
F & 1.1$\pm$0.2 & 2.8$\pm$1.8 & 1.7$\pm$0.2 &$>$16&$\sim$4.1 & 2.4$\pm$0.7\\
G & 0.9$\pm$0.1 &1.5$\pm$0.7 & $\sim$2.2 &$>$11&$\sim$3.8 & 1.1$\pm$0.4\\
H & 1.0$\pm$0.1 &\nodata & \nodata &$>$4.4&$>$1.0 & $<$2.4\\
\enddata
\tablecomments{The resolution of the measurements are set by the 
lowest resolution input maps. The HCN(1-0) is from 
\citet[][]{MTH08} and the CS(2-1) from \citep[][]{KNSS08}.  In determining 
the CS/C$^{34}$S ratio it is assumed that CS is spatially 
completely resolved by the \citet[][]{KNSS08} map.  If this assumption is incorrect 
true ratios will be slightly lower than those quoted.  Uncertainties reflect absolute 
flux calibration except for the C$_{2}$H ratios which are based on 
signal-to-noise.  Upper limits are 1$\sigma$.}
\end{deluxetable}

\begin{deluxetable}{lcccccccccccccc} 
\tabletypesize{\scriptsize}
\tablenum{7} 
\tablewidth{0pt} 
\tablecaption{PCA Correlation Matrix\label{pcacorT}} 
\tablehead{ 
\colhead{Maps}  
&\colhead{ $^{13}$CO}
&\colhead{3MM} 
&\colhead{ C$_{2}$H($\frac{3}{2}$)}  
& \colhead{ HCN}  
& \colhead{ HCO$^{+}$}
& \colhead{ HNC}  
&\colhead{ CH$_{3}$OH} 
& \colhead{ HNCO}  
& \colhead{N$_{2}$H$^{+}$}
&\colhead{ HC$_{3}$N}  
& \colhead{ C$^{34}$S}  
& \colhead{ SiO}  
&\colhead{C$_{2}$H($\frac{1}{2}$)}  
}  
\startdata 
$^{13}$CO&1.00 &&&&&&&&&&&& \\
3MM&0.88 &1.00 &&&&&&&&&&& \\
C$_{2}$H($\frac{3}{2}$)& 0.85 & 0.77 & 1.00&&&&&&&&&& \\
HCN& 0.93 &0.92& 0.84 &1.00&&&&&&&&\\
HCO$^{+}$& 0.89 &0.88 &0.81 &0.98 &1.00&&&&&&&&\\
 HNC& 0.95 &0.87 &0.84 &0.97 &0.94 &1.00&&&&&&&\\
CH$_{3}$OH& 0.68 &0.47 &0.49 &0.43 &0.34 &0.51 &1.00&&&&&&\\
 HNCO& 0.52 &0.27 &0.32 &0.21 &0.12 &0.33 &0.86 &1.00&&&&&\\
N$_{2}$H$^{+}$ &0.86 &0.71 &0.67 &0.77& 0.72& 0.82& 0.71& 0.58 &1.00&&&&\\
HC$_{3}$N& 0.54 &0.53 &0.41& 0.62 &0.60 &0.62& 0.32& 0.16 &0.63 &1.00&&&\\
 C$^{34}$S& 0.20 &0.24 &0.17 &0.18& 0.20& 0.19& 0.14& 0.095& 0.20& 0.15& 1.00&&\\
SiO& 0.38 &0.23& 0.34 &0.34 &0.30 &0.38 &0.35 &0.30 &0.35 &0.31&-0.023 &1.00&\\
C$_{2}$H($\frac{1}{2}$)& 0.64 &0.60& 0.70 &0.66 &0.63 &0.63 &0.27 &0.18 &0.45 &0.24 &0.008 &0.32 &1.00 \\
\enddata 
\tablenotetext{a}{Data from \citep[][]{MTH08}.}
\end{deluxetable} 
 
\begin{deluxetable}{lccccccc} 
\footnotesize
\tablenum{8} 
\tablewidth{0pt} 
\tablecaption{PCA Eigenvectors \label{pcaT}} 
\tablehead{ 
\colhead{PCA Comp.}  
&\colhead{1}
&\colhead{2}
&\colhead{3} 
&\colhead{4}  
&\colhead{5}  
&\colhead{6} 
&\colhead{7}
}  
\startdata 
$^{13}$CO& 0.35&  -0.045&  -0.0043 &0.11 & -0.042 &0.15  &-0.064 \\
3MM&0.32 &0.15  & -0.10 &0.12 & -0.086 &0.19 & -0.50 \\
C$_{2}$H($\frac{3}{2}$)& 0.31& 0.097 & 0.086 & 0.24 & 0.12 & 0.020 & 0.81 \\
HCN& 0.34 & 0.21 & 0.00 &  -0.012 &  -0.057 & 0.16 &  -0.086 \\
HCO$^{+}$&  0.33 &0.28  &-0.029  &-0.024  &-0.055  &0.21  & -0.038 \\
HNC& 0.35 &0.11 &0.0015  &-0.031  &-0.054 &0.16  &0.038 \\
CH$_{3}$OH&  0.23 & -0.55 & -0.03 &0.10  &-0.091 &0.0061  &-0.031 \\
HNCO&  0.17  &-0.65  &-0.01 &0.17  &-0.050  &-0.10  &-0.061 \\
N$_{2}$H$^{+}$ &0.32  &-0.19 & -0.089 & -0.12  &-0.20  &-0.061 &0.070 \\
HC$_{3}$N& 0.23 &0.067  &-0.13  &-0.68  &-0.29  &-0.52 &0.086 \\
C$^{34}$S&  0.080&  -0.0027  &-0.81 &0.0092 &0.56  &-0.11  &-0.0031 \\
SiO&  0.16  &-0.19 &0.43  &-0.51 &0.66 &0.23  &-0.074 \\
C$_{2}$H($\frac{1}{2}$)&  0.24 &0.19 &0.33 &0.37 &0.29  &-0.70 & -0.23 \\
\hline
Egnv. \% &59&13&8.3&6.6&5.6&2.7&1.5 \\
\enddata 
\end{deluxetable}

\end{document}